# Prospects for Improving the Intrinsic and Extrinsic Properties of Magnesium Diboride Superconducting Strands

E.W. Collings, M. D. Sumption, M. Bhatia* M. A. Susner, and S. D. Bohnenstiehl,

Laboratories for Applied Superconductivity and Magnetism,

Department of Materials Science Engineering, The Ohio State University

477 Watts Hall, 2041 College Rd, Columbus OH 43210

*Now at the Texas Accelerator Center, The Woodlands, TX 77381

## Abstract

The magnetic and transport properties of $MgB_2$ films represent performance goals yet to be attained by powder-processed bulk samples and conductors. Carbon-doped films have exhibited upper critical fields, $\mu_0 H_{c2}$, as high as 60 T and a possible upper limit of more than twice this value has been predicted. Very high critical current densities, $J_c$, have also been measured in films, e.g. 25 MA/cm$^2$ in self field and 7 kA/cm$^2$ in 15 T. Such performance limits are still out of the reach of even the best $MgB_2$ magnet wire. In discussing the present status and prospects for improving the performance of powder-based wire we focus attention on (1) the *intrinsic* (intragrain) superconducting properties of $MgB_2$ -- $H_{c2}$ and flux pinning, (2) factors that control the efficiency with which current is transported from grain-to-grain in the conductor, an *extrinsic* (intergrain) property. With regard to Item-(1), the role of dopants in $H_{c2}$ enhancement is discussed and examples presented. On the other hand their roles in increasing $J_c$, both via $H_{c2}$ enhancement as well as direct fluxoid/pining-center interaction, are discussed and a comprehensive survey of $H_{c2}$ dopants and flux-pinning additives is presented. Dopant selection, chemistry, methods of introduction (inclusion), and homogeneity of distribution (via the rounding of the superconducting electronic specific heat transition) are considered. Current transport through the powder-processed wire (an extrinsic property) is partially blocked by the inherent granularity of the material itself and the chemical or other properties of the intergrain surfaces. Overall porosity, including reduced density and intergranular blocking, is quantified in terms of the measured temperature dependence of the normal-state resistivity compared to that of a clean single crystal. Several experimental results are presented in terms of percent effective cross-sectional area for current transport. These and other such results indicate that in many cases less than 15% of the conductor's cross sectional area is able to carry transport current. It is pointed out that densification in association with the elimination of grain-boundary blocking phases would yield five-to ten-fold increases in $J_c$ in relevant regimes, enabling the performance of $MgB_2$ in selected applications to compete with that of $Nb_3Sn$.

## Outline

## 1.0. Introduction

The initial announcement of superconductivity below a $T_c$ of about 40 K in $MgB_2$ by Akimitsu in 2001[1] triggered extensive research on its preparation in film, bulk, and wire forms as well as its electrical, magnetic, and mechanical properties. Already by June of that year more than 250 research papers had been published and by November 2001 $MgB_2$ was the subject of a comprehensive review by Buzea and Yamashita [2]. With upper critical fields, $H_{c20}$, no higher than about 18 T in single crystals, bulks, and wires, and with critical current densities, $J_c$, mostly less than about $10^4$ A/cm$^2$ at 5 K, 5 T, $MgB_2$ tended to be of scientific rather than practical interest. But it was the report of very high $H_{c2}$s in alloyed ("dirty") thin films, accompanied by promisingly high $J_c$s by Eom et al. in May 2001 [3], along with the subsequent reports of $H_{c20}$s of more than 70 T by Braccini et al [4] in 2005, that alerted the community to the possibility of wide-spread practical applications if such properties could be transferred to tapes and wires on a large scale. Subsequent years have seen the establishment of commercial companies dedicated to the manufacture of bulk $MgB_2$ objects and long lengths of $MgB_2$ tapes and wires. At the same time steady improvements in $H_{c2}$ and $J_c$ have been reported. As a result $MgB_2$ is on the threshold of widespread practical application, an important example being magnets for magnetic resonance imaging. But to take the next step, whether we are interested in high field 4 K operation, or in the more energy efficient 20 K temperatures regimes well out of the range of the present NbTi and $Nb_3Sn$ wires, further improvement of the in-field $J_c$ is needed. In addressing this issue we must recognize two classes of properties (what might be termed "intrinsic" and "extrinsic", respectively) that are in urgent need of fundamental research. Below, we summarize some of the key issues that need to be addressed.

### *1.1. Intrinsic Properties*

The essential intrinsic (i.e.intragranular) properties of polycrystalline $MgB_2$ are $H_{c2}$ and flux pinning (or, alternatively, intra-grain $J_c$). Increases in the former are always welcome, but the already measured $H_{c20}$s of more than 33 T [5] are usefully high provided they are homogenous. That this is not yet the case is all too evident in experimental results to-date. We must also consider the anisotropy of the superconducting properties, as manifested particularly in the $H_{c2}$ values for field along the basal planes as compared to those along the *c*-axis. This anisotropy can be addressed at an intrinsic level by dopant selection, as some dopants modify this intrinsic anisotropy. Anisotropy also has implications for grain-to-grain current flow in untextured polycrystalline materials, but this is better addressed as a kind of effective connectivity within the extrinsic group of properties.

Beyond this, there must be a satisfactorily high level of flux pinning. Such is not the case at present; flux pinning centers yielding high bulk pinning forces, $F_p$, with maxima located at fields much higher than the existing 4-5 T at 4 K (i.e., at about 20% of the irreversibility field, $H_{irr}$) are preferred. Extensive research has already been carried out on the effects of dopant additions on $H_{c2}$ and $J_c$. Indeed, numerous metallic elements, borides, carbides, oxides, and even organic compounds have been added. This research needs to be continued on a rational scientific basis and redirected if $MgB_2$ is to operate successfully at higher fields and temperatures. Necessary to generate improved properties is the presence of suitably chosen *combinations* of dopants (elements in substitutional solid solution) plus nano-size flux pinning additions (e.g. [6]). However, as will be argued below, the two most crucial improvements needed to significantly enhance $MgB_2$ performance are: (i) homogenous dopant induced $H_{c2}$ enhancement, and (ii) large increases in connectivity (the removal of inter-grain resistive boundaries).

Texturing (or anisotropy reduction) and flux pinning, while important, will not influence final transport $J_c$ as strongly as those properties. While homogeneous doping and $H_{c2}$ enhancement can be considered intrinsic properties, connectivity is an inherently extrinsic property.

### *1.2. Extrinsic Properties- Connectivity and Porosity Issues*

In addition to intrinsic-property improvement substantial across-the-board increases in $J_c$ will accompany increases in the superconductor-material's effective cross-sectional area for the conduction of transport current. The possibility of "cross-sectional deficiency" was first recognized by Rowell [7][8] who noted that the normal-state electrical resistivities of some polycrystalline thin film samples of $MgB_2$ could be as high as 500 μΩcm – well beyond the regime of metallic conductivity. Not only normal-current flow, but supercurrent flow is also expected to be reduced by this cross-sectional deficiency, removal of which is estimated to increase the $J_c$ of a given polycrystalline sample by factors of five-to ten. Although clean grain boundaries as such do not impede supercurrent flow [9], the observed intergranular partial coatings of insulating MgO or amorphous $BO_x$ [10] will certainly do so. Formation of the coating, an interesting problem in solid state chemistry, is an area of active study. Research is focussing on the possibility of reducing or eliminating it [11] leading to improved intergranular connectivity.

Magnesium diboride wires are made by powder-in-tube processes, either *ex-situ* in which ready-made $MgB_2$ powder is packed into a tubular jacket and drawn down or *in-situ* in which the jacketed elemental Mg and B ingredients are reacted together at final wire size. Although the ex-situ wire is subject to the porosity expected of any powder-processed body, the in-situ wire filling is made even more porous by the voids remaining from the prior Mg particles after the $Mg+2B \rightarrow MgB_2$ reaction. Various consolidation techniques are beginning to be introduced in attempts to reduce porosity in bulk samples and wires. Different reaction routes to $MgB_2$ formation that will lead directly to reduced porosity are also being studied [12].

## 2.0. Intrinsic Properties of $MgB_2$ Conductors

The very high $H_{c2}$s reported for dirty films ([3] and subsequently [4]) which alerted the community to the possibility of high field and other practical applications, continue to serve as useful references against which the $H_{c2}$s of evolving $MgB_2$ conductors can be compared. Likewise the high $J_c$s of dirty films can be seen as goals, bearing in mind that although films are not immune to connectivity problems, at least they are not porous.

### *2.1. Upper Critical Fields and Irreversibility Fields and their Enhancement by Dopant-Induced Impurity Scattering*

***2.1.1. Theory:*** For dirty, non-paramagnetically limited, single-band, low-temperature superconductors the following well known expression for zero-K upper critical field, $H_{c20}$ can be obtained by starting with Maki's dirty-limit $H_{c20}$ and inserting an expression for the BCS zero-K thermodynamic critical field, $H_{c0}$, thus:

$$\rho_0 = \frac{3.27x10^{-5}}{\gamma}\left(\frac{H_{c20}}{T_c}\right) = \frac{2.23x10^{-5}}{\gamma}\left(\frac{-dH_{c2}}{dT}\right)_{T_c} \quad (1)$$

where $\rho_0$ is the impurity resistivity and $\gamma$ the electronic specific heat coefficient. Equation (1) indicates that the value of $H_{c2}(T)$ at 0 K and its slope at $T_c$ are proportional to $\rho_0$ (alternatively, inversely proportional to the charge-carrier diffusivity, where $1/\rho_0 = e^2 N_F D$). Increases in $H_{c20}$ would be expected to follow directly from the introduction of charge-carrier-scattering impurities into the superconductor's lattice structure - a rule that has guided the design of low temperature superconductors over the years. But this simple rule is no longer obeyed for $MgB_2$, for which there is no direct relationship between $H_{c20}$ and $\rho_0$ [13].

In the compound $MgB_2$ the presence of Mg stabilizes the B sub-lattice in the form a honeycomb-like stack of hexagonal networks, essentially a new allotrope of B. The B honeycomb dominates the electronic structure which can be thought of as deriving from σ bonding within the B plane (leading to a "σ band" ) and π bonding both in and out of the plane (leading to a "π band") [14]. The bands have their individual Fermi densities-of-state, superconductive energy gaps, and scattering relation times, $D_\sigma$ and $D_\pi$, respectively [15].

With reference to a copious body of experimental upper critical field results comparable to that of Figure 1, Gurevich et al [16][17] [18] have been able to illustrate and explain $MgB_2$'s unusual critical-field temperature dependence in terms of the electronic diffusivities ($D_\pi$ and $D_\sigma$) associated with its two-band conductivity:

$$\sigma = e^2 (N_{F\pi} D_\pi + N_{F\sigma} D_\sigma) \tag{2}$$

The introduction of two diffusivities results in pronounced departures from the single-band predictions; in particular, $H_{c2nearTc}$, and $H_{c20}$ respond individually to $D_\pi$ and $D_\sigma$. Thus, according to Gurevich et al [16] and to a first approximation:

$$H_{c2nearTc} \propto \frac{1}{\lambda_1 D_\sigma + \lambda_2 D_\pi} \tag{3}$$

(in which the $\lambda$s are appropriate functions of the intraband and interband electron-phonon coupling constants [16], see also Section 4.1.2) and

$$H_{c20} \propto \frac{1}{\sqrt{D_\sigma D_\pi}} \tag{4}$$

Simply stated, near the critical temperature $H_{c2nearTc}$ varies inversely as the weighted arithmetic mean of $D_\sigma$ and $D_\pi$, while $H_{c2near0K}$ ($H_{c20}$) varies inversely as their geometric mean. Thus while increasing impurity scattering is beneficial to $H_{c2}$ over the entire temperature range for both single-band and double-band superconductors, in the latter case (and recognizing the independence of $D_\pi$ and $D_\sigma$) it allows $H_{c20}$ to diverge to very large values in response to decreases in either $D_\sigma$ or $D_\pi$.

***$H_{c2}$ at "High Temperatures":*** Suppose, as has turned out to be the case for dirty $MgB_2$ films, $D_\pi \ll D_\sigma$ (i.e., π scattering much stronger than σ scattering) Equation (3) approximates to

$$H_{c2nearTc} \propto \frac{1}{\lambda_1 D_\sigma} \tag{5}$$

$H_{c2nearTc}$, which is in general controlled by the **greater** of the two diffusivities, is in this case controlled by $D_\sigma$. Further changes in $H_{c2nearTc}$ would be expected to follow a manipulation of $D_\sigma$.

***$H_{c2}$ at "Low Temperatures":*** Referring to Equation (4) in which $D_\sigma$ and $D_\pi$ are tied together as a product (even though $D_\pi$ is the "dominant" component), it would seem that manipulation of either of them would influence $H_{c20}$. But examination of the $H_{c20}$ relationship in more detail reveals that for very different diffusivities

$$H_{c20} \propto \frac{1}{D_\sigma} \qquad D_\sigma << D_\pi \tag{6}$$

$$H_{c20} \propto \frac{1}{D_\pi} \qquad D_\pi << D_\sigma \tag{7}$$

so the $H_{c20}$ is controlled by the smaller of the two diffusivities – e.g., the $D_\pi$ in dirty $MgB_2$ films. As pointed out by Gurevich [18] $H_{c20}$s close to the BCS paramagnetic limit of $H_p$ = 65 T (see Figure 1 [19]) are already being attained in $MgB_2$. Furthermore, given that BCS underestimates $H_p$ by the factor $(1+\lambda)$ where $\lambda$ is some global electron-phonon coupling constant, values as high as 130 T may in principal be achievable. Based on the spectacular thin film results Gurevich et al. [16] pointed out that the use of $MgB_2$ will open a "new domain of application for superconducting magnets", one that is outside the reach of NbTi and to some extent $Nb_3Sn$, implying that in general enhancements to the $H_{c2}$ will follow the judicious introduction of scattering impurities.

***2.1.2. Selection of Dopants for $H_{c2}$ Optimization:***: It has been suggested that: (i) macroscopic particles that contribute to lattice distortion enhance both π and σ scattering, (ii) disorder on the Mg sublattice (e.g. by Al addition) can increase the π scattering, (iii) oxygen when substituting for B was expected to provide strong σ scattering; so also was C [16]. By now a multitude of papers have described the benefits of adding C in one form or another to the $MgB_2$ lattice. Carbon has been added in elemental form (microspheres, nanotubes, nanodiamond) or by way of C-containing compounds such as SiC, TiC, $B_4C$, organic compounds, and so on (see below). The work has been largely semi-empirical in nature albeit related to and guided by the principles outlined above.

Although C when doped into the B lattice might be expected to increase σ scattering (as predicted by recent electronic-structure calculations [20]), Angst et al [21] and more recently Gurevich [18] have offered experimental evidence in support of a π-scattering mechanism. These authors agree that the upward curvature of the low-temperature $H_{c,field\ parallel}$, or at least its lack of negative curvature, see Figure 1, points to the dominance of π-band scattering. Although dopant induced buckling of the *a-b* (Mg) planes has been suggested as a mechanism for the enhanced π scattering [18][21] the latter authors have concluded that it is in fact an intrinsic result of C substitution but for reasons that are not fully understood [18][21].

In selecting dopants, reaction chemistry is an important consideration. The success of SiC as a dopant for C [22] relies on the fact that, during in-situ processing, the Mg decomposes SiC and releases atomic C before the onset of the $Mg+2B \rightarrow MgB_2$ reaction. This is clearly demonstrated in the results of differential scanning calorimetry (DSC), Figure 2. All DSC data depicted there were collected with a TA Instruments DSC2920 under argon flow, using an Fe pan. Three different runs were made: (1) The first with Mg only (solid line) exhibits the endothermic peak associated with Mg melting at about 650°C. (2) A run made just with mixed SiC and Mg powders (the dashed line) evinces the reduction of SiC by Mg at 524°C again followed by Mg melting at 650°C. (3) The third curve (dash-dot) shows the response of a sample of mixed Mg and amorphous B (99+% pure) powders. It depicts the endothermic surface reaction described in [23] as well as the onset of $MgB_2$ formation at 615°C below the Mg's 650°C melting point [23].

Since dopant concentrations are generally small, attention must be given to ensuring that they are uniformly distributed and/or incorporated into the lattice. Thinking along these lines, Ribiero et al [24] selected $B_4C$ as a source of C doping reckoning that the C and the B were already mixed on an atomic-length scale. A similar strategy, and other considerations, also guided the selection of the diborides $TiB_2$, $ZrB_2$, and $NbB_2$ as potential $H_{c2}$-enhancing dopants [5]. Of these, $ZrB_2$ was the most potent, and with an $H_{c2,4K}$ of 29 T, became the subject of more detailed study [25], Figure 3. Two bulk samples were made, the first with binary stoichiometric $MgB_2$ and the second with 7.5 mol% $ZrB_2$ addition. They were prepared by the in-situ reaction of a stoichiometric mixture of 99.9% pure Mg and amorphous B (99+%) powders after SPEX milling and compacting into pellets. Critical field measurements were performed resistively at 4 K, and magnetically at higher temperatures [25]. In Figure 3, the filled triangles give the $H_{irr}$ vs $T$ for binary $MgB_2$, the filled circles give the response of the $ZrB_2$-doped sample, and the open triangles represent the percentage increase in $H_{irr}$ resulting from that doping. Based on XRD measurements, which showed increases in both the *c*- and *a*-lattice parameters, we deduced that Zr was substituting on the Mg site. Thus, we might expect the dopant to be effective primarily at lower temperatures (assuming $\pi$ band scattering). It is interesting to note that the increase for $ZrB_2$ [25] is not very different from that contributed by SiC doping [22]. Additionally, even with substitutional doping the possibility of unreacted residues cannot be ignored. Although they were not present as a result of $ZrB_2$ doping, they were observed with $NbB_2$ doping [25] and are known to be present after SiC doping [22].

Of course, the unreacted residue is not necessarily inert, and may contribute flux pinning centers, or indirectly increase either $\pi$ and $\sigma$ scattering via lattice distortion and in that way increase $H_{irr}$. However, although numerous dopants (including metallic elements, borides, carbides, oxides, and organic compounds) have been added to $MgB_2$ little direct attention has been paid to the possible existences of such dual doping mechanisms.

### *2.2. Critical Current Density and its Enhancement by Doping and Flux Pinning*

***2.2.1. Doping in General:*** In considering the $J_c$s attainable in $MgB_2$ bulk samples we turn again, as a first basis of comparison, to the results of thin films measurements, with reference to the work of Eom et al [3] and many subsequence researchers. Figure 4 displays the $J_c$s of $MgB_2$ films made by various techniques -- e-beam evaporation, thermal co-evaporation, hybrid physical-chemical vapor deposition, and pulsed-laser deposition (PLD). There are two versions of PLD – "in-situ" in which the $MgB_2$ is directly deposited and "ex-situ" in which a PLD-deposited B layer is exposed to Mg vapor, although in practice, some post-deposition of Mg seems to be needed even in the "in-situ" case.

These thin-film "reference $J_c$s" are compared to those of commercially processed binary $MgB_2$ powder-in-tube (PIT) wires of varying filament counts in Figure 5. These strands were all made via the in-situ route by Hyper Tech Inc. from mixed Mg powder (10% excess Mg) and 99% pure amorphous B powder. They typically achieved 4 K $J_c$s of $10^5$ A/cm$^2$ at 4.5-5 T, and $10^4$ A/cm$^2$ at 8 T, about an order of magnitude below the best thin film data. We go on to show some incremental improvements in $J_c$ that result from: (i) the introduction of even more Mg and (ii) the doping of the excess-Mg strand with SiC particles, Figure 6. For this sample set the best results ($10^5$A/cm$^2$ and $10^4$ A/cm$^2$ at 6 T and 11.5 T, respectively) are found when the excess-Mg level is raised to 15% and SiC is added. Still further improvements are depicted in Figure 7 which emphasizes the effectiveness of the C-containing dopants e.g. maleic anhydride, SiC, and C itself when appropriately introduced. On the best of such samples the $10^4$ A/cm$^2$ $J_c$ is obtained at a field as high as 13 T. Although this result could be said to be about a factor of three off from the best thin film data reported at these fields, there is in fact a paucity of data in this regime, some of the higher performing films having been measured only at lower fields.

Over the years some 40-odd dopant species have been introduced into $MgB_2$, Table 1. It is possible to discern a few guiding principles that may have governed their selection. The inclusion of the metallic elements and metal oxides were clearly attempts to introduce normal flux-pinning centers. The borides, nitrides and C-containing dopants were added to increase $H_{c2}$ and hence to increase the effectiveness of already present pinning centers as described below. Some additions likely played a dual role – decomposing to partially substitute into the $MgB_2$ lattice while leaving a flux-pinning particulate residue.

In concluding this section on $J_c$ enhancement it is important to note that: (i) The primary purpose of a few of the compositional modifications and/or dopant additions (e.g., $CaC_6$) is to reduce grain-boundary contamination and improve connectivity, the subject Section 3 of this paper. (ii) Although useful as interim goals, the thin-film $J_c$s of Figures 4-7 are not the ultimate "gold standards"; essential to the Rowell connectivity analysis [7][8] was the fact that even polycrystalline films are subject to intergrain transport degradation. Even so, it is important to make and then measure films out to high fields at 4 K to explore the potential of $MgB_2$ in this field regime.

***2.2.2. Flux Pinning and Doping Mechanisms:*** Apart from pure single crystals all $MgB_2$ samples – films, bulks, and wires – are polycrystalline in the as-prepared state and hence tend to be moderately well grain-boundary pinned. The introduction of dopants can have one or more effects which may combine to increase the bulk pinning strength: (1) by increasing the crystal's $H_{c2}$ and $H_{irr}$; (2) by forming a wide distribution of point pinning centers each described by an elementary pinning force $f_p$; (3) by producing localized lattice strains which also contribute to flux pinning. The magnitude and behaviour of $J_c$ in an applied field $H$ is described by a bulk pinning function, $F_p$, which represents an appropriate **summation** of the elementary pinning forces, $f_p$ [92-94]. Summation leads to expressions for the bulk pinning force density originally written in the form:

$$F_p = H_{c2}^{m}.f(h) \qquad \text{where } h = H/H_{c2} \tag{8}$$

Recognizing that for $MgB_2$ with an $H_{irr}$ (the actual field at which $F_p$ vanishes) that is significantly less than $H_{c2}$ we need to define $h = H/H_{irr}$. Although part of the prefactor in the expression for $F_p$ derives from the field-normalization step, it is significant that another component comes from a condensation energy. Its presence indicates that dopants added to increase $H_{c2}$, although not pinning centers themselves, serve to increase the measured $F_p$ of already-pinned samples.

The bulk pinning force field dependence, which generally rises to a maximum at some $h = h_{max}$, may be expressed in normalized form $F_p(h)/F_p(h_{max})$. This, for a given type of pinning center, is independent of temperature, a property known as ***scaling.*** The quotient $F_p/F_{p,max}$ is of the general form

$$f(h) \propto h^p(1-h)^q \tag{9}$$

As pointed out by Dew-Hughes [92] various classes or types of pinning by "normal" (non-superconducting) objects can be distinguished by the values assigned to the index $p$ (with $q = 2$ for all normal pins), thus

| | | | |
|---|---|---|---|
| $h^0(1-h)^2$ | $\rightarrow$ | normal volume pinning | $d < a,b,c$ |
| $h^{1/2}(1-h)^2$ | $\rightarrow$ | normal surface (grain boundary) pinning | $d < a,b$ |
| $h^1(1-h)^2$ | $\rightarrow$ | normal point pinning | $d > a,b,c$ |

where *a, b, c,* characterize the dimensions of the pin and $d$ is the fluxoid spacing; $d^2 = 1.15(\phi_0/B)$. Normal surface and point pinning can also be distinguished in terms of the corresponding values of $h_{max}$, viz. 0.2 and 0.33, respectively. Examples of the above normal pinning functions together with pinning-force data for a SiC-doped $MgB_2$ sample are given in Figure 8. This figure displays pronounced departures from scaling. The generally expected characteristic of grain-boundary pinning is seen for temperatures between 10-20 K. In the 20-30 K range, it appears that point pinning is beginning to dominate. At 20 K, we seem to have an intermediate case. Our normalizing field, $H_{irr}$, was based on a 100 $A/cm^2$ transport criterion. The lack of fit to the curves at higher temperatures can be expected from a distribution of $H_{irr}$ values (and underlying $H_{c2}$ values) resulting from sample inhomogeneity. Evidence for this can be seen in the strong sensitivity of $H_{irr}$ to measurement criteria, as well as in the heat capacity behaviour discussed below.

We have been referring above to normal pins -- i.e. non-superconducting pinning particles or regimes. Objects with $\kappa$-values differing markedly from the matrix pin by the so-called "$\Delta\kappa$" mechanism which, according to Dew-Hughes [92] is characterized by a $q$-index of 1. Based on this criterion, none of the centers we have encountered in $MgB_2$ have exhibited $\Delta\kappa$ pinning.

### *2.3 Electronic Specific Heat and Compositional Inhomogeneity*

In selecting and introducing potential $H_{c2}$- enhancing dopants it is important to be alert to the likelihood of compositional inhomogeneity in the resulting doped sample. In adding small fractions of dopant to a sample it is difficult to guarantee that the starting particles are uniformly dispersed. Even if they are, during reaction heat treatment the existence of finite diffusion distances will essentially guarantee a certain levels of inhomogeneity. Such will manifest themselves in terms of deviations from expected pinning curves -- difficult to de-convolute and separate from the effects of mixed pinning centers. However, analysis of the electronic specific heat jump of a superconducting material in the vicinity of its $T_c$ can provide a sensitive quantitative measure of its degree of compositional homogeneity. The total low temperature specific heat of a metal ($C_p$) is the sum of a lattice (phononic) component , $C_l$, given by:

$$C_l = 9Nk\left(\frac{T}{\Theta_D}\right)^3 \int_0^{\Theta_D/T} \frac{x^4 e^x}{\left(e^x - 1\right)^2}\,dx \qquad (10)$$

in which $N$ = Avogadro's number, $k$ = Boltzmann's constant (hence $Nk$ = 8.31 J/mole.K) and an electronic component, $C_e$, given by $\gamma T$ for a normal metal and $\gamma[T_c .a.exp(-b/T)]$ for a superconductor below its transition temperature where $a$ and $b$ are metal-specific constants. At temperatures around the $T_c$ of $MgB_2$ $C_e$ is negligible compared to $C_l$ (cf. Figures 9(a) and (b)). This has two consequences: (1) The specific heat data can be fitted to Equation (10) enabling a calorimetric-$\theta_D$ to be extracted. (2) A special technique must be invoked to expose the electronic component. On the assumption that the latter is magnetic field independent the superconductive $C_e$ and its jump at $T_c$ can be revealed and expanded after the in-field-measured specific heat is subtracted from the zero-field data. For this purpose a PPMS machine was used to take measurements in a magnetic field of 9 T over the temperature range 20-40 K. Figure 9(a) shows the results of $C_p$ measurements on bulk samples of (i) binary stoichiometric $MgB_2$, (ii) $MgB_2$ doped with 10% SiC (200 nm powder size), (iii) $MgB_2$ doped with 7.5% $TiB_2$ (<44 μm before milling) prepared as described in [5]. The curves, are similar in appearance, responding primarily to differences in Debye temperatures (see below) and with $C_e$ jumps at $T_c$ unresolved on the scale of the measurement. Figure 9(b), provides a detailed view of the electronic specific heat and its transition at $T_c$. A close look at the data reveals onset temperatures that decrease roughly in the sequence 38 K (MB700), 36.5 K (MBTi800), 35.5 K (MBSiC700) in agreement with the magnetically measured onsets [5] and in accord the results of Wälti et al who recorded a $T_c$ of 37.5 K in their compacted pellet. Decreases in $T_c$ below the nominal value of ~39 K are attributed to the presence of impurities [95,96] and in our case to the presence of dopants. The breadths of the superconducting jumps in Figure 9(b) contain information about the distribution of transition temperatures. No attempt is made here to fit the data to such a distribution although it is possible to do so using standard techniques [97-99], nevertheless the presence of compositional inhomogeneities in the heavily doped MBTi700 and MBSiC700 is at least revealed qualitatively in the figure.

## 3.0. Extrinsic Properties of Polycrystalline $MgB_2$

### *3.1. Porosity*

Randomly packed similar spheres will occupy space to a volume density of about 65% (although spheres of mixed sizes will fill the space more efficiently) for which reason compacted-powder pellets and the cores of PIT wires are inherently porous, e.g. Figure 10. Thus the packing densities of ex-situ-processed bulks and wires, based geometrically on the packing densities of powders, will be around 65%. Those of in-situ-processed bulks and wires are complicated by shrinkage associated with the $Mg+2B \rightarrow MgB_2$ reaction, but can be estimated using the data from Table 2. Our limiting estimates are based on two model scenarios.

***Model-(1):*** *One mole of $MgB_2$ produced in-situ within a fixed-volume enclosure from a mixture of equi-sized Mg and B particles packed to a volume density-factor $P_i$.*

From Table 2 the volume ratio [Final ($MgB_2$)/Initial (Mg+2B)] = 17.2/23.0 = 0.75. It follows that if the initial packing density is $P_i$ then the final density is $P_{f1} = 0.75P_i$. For example an initial mixed-powder close-packing of 65% leads to an $MgB_2$ powder density of 49%.

***Model-(2):*** *One mole of $MgB_2$ produced in-situ within a fixed-volume enclosure from coarse Mg particles imbedded in fine B powder itself packed to a volume density-factor $P_i$ – a kind of "micro-infiltration method"[100] more closely resembling reality.*

From Table 2, 13.74 cm$^3$ of Mg is required to produce 1 mol (= 17.21 cm$^3$) of $MgB_2$. The Mg pieces are imbedded in 9.25/$P_i$ cm$^3$ of fine B powder (as in the Giunchi method, e.g. [100], but on a smaller scale). The overall packing density after reaction is therefore $P_{f2}$ = 17.21/[13.74+(9.25/$P_i$)]. In this case, for example, a 65% dense B-powder environment would lead to an overall $MgB_2$ powder density of 62%. Taken together, these results as depicted in Figure 11 indicate that conventionally in-situ-processed bulks and PIT wires can be expected have packing densities in the range of 50-60% depending on the relative sizes of the Mg and B particles or powders.

Porosity impedes current flow and prevents the $J_c$ of the $MgB_2$ body from attaining its intrinsic intragranular value. Furthermore, transport current flow is impeded by intergranular partial coatings, e.g. insulating MgO or amorphous $BO_x$ [10]. The combined impact of porosity and current-blocking phases, illustrated in Figure 12, is to reduce the conductor's effective cross-sectional area for current flow by the inverse of $F$, the "Rowell factor" (1/$F$ is less than 1)..

The possibility of "cross-sectional deficiency" was first recognized by Rowell [7][8] who noted that the normal-state electrical resistivities of some polycrystalline thin film samples of $MgB_2$ could be as high as 500 μΩcm – well beyond the regime of metallic conductivity. Supercurrent flow is also expected to respond to respond to this same cross-sectional deficiency, which causes the measured extrinsic $J_c$ of a polycrystalline sample to be very much less than its intrinsic value.

### *3.2. General Connectivity*

Since the effect of general porosity can be regarded as a geometrical phenomenon it is postulated that normal-state resistivity can be used as a measure of effective cross-sectional area for transport in both the normal and superconducting states. Our connectivity analysis, which uses a modification of the Rowell approach [7,8], requires a detailed resistivity temperature dependence measurement but in response yields some additional information relevant to the sample's superconducting properties. We begin with an expression for the measured total resistivity:

$$\rho_m(T) = \rho_{0,m} + \rho_{i,m}(T) = F[\rho_0 + \rho_i(T)] \tag{10}$$

in which $\rho_{0,m}$ is the measured residual resistivity and $\rho_{i,m}(T)$ is the measured ideal (phononic), component. Finally, on the right-hand-side the terms within the brackets, $\rho_0$ and $\rho_i(T)$, are the intragrain residual- and ideal (phononic) resistivities, respectively, and the prefactor, $F > 1$, represents the observed resistivity enhancement due to the reduced effective transport cross-sectional area. Our approach consists in replacing the previously assumed universal $\rho_i(T)$ with the Bloch-Grüneisen function, thus:

$$\rho_m(T) = F\left[\rho_0 + \frac{k}{\theta_D}\left(\frac{T}{\theta_D}\right)^5 \int_0^{\theta_D/T} \frac{e^z z^5 dz}{\left(e^z - 1\right)^2}\right] \tag{11}$$

in which $\theta_D$ is the Debye temperature and $k$ is a "materials constant" which we evaluate by fitting Equation (11) to a published set of single crystal data for which $F \equiv 1$. Fits of Equation (11) to the measured resistivity data then yield the following information: (1) the $F$-factor (and hence the percent connectivity, $100/F$) which enables the true intragrain $J_c$ to be determined from the results of an accompanying $J_c$ measurement; (2) the intragrain residual resistivity, $\rho_0$; (3) a resistive Debye temperature, $\theta_R$, available for comparison to its calorimetrically measured counterpart, $\theta_D$, see below.

By way of example we draw on the results of resistivity measurement on three pellets measured above for heat capacity; one binary stoichiometric sample, one doped with 7.5 % $TiB_2$, and another doped with 10 % SiC. The results shown in Figure 13 after fitting to Equation (11) yielded the values of $\rho_0$, $\theta_R$, $F$, and percent connectivity (100/$F$) that are listed in Table 3.

Connectivity is the most practically significant result of the resistivity experiment. In Table 3 we note that the connectivities of the polycrystalline grains of the unalloyed and doped $MgB_2$ pellets vary from 11 to 23%. If these materials are representative of PIT wire cores, as no doubt they are, we can expect substantial increases in $J_c$ e.g. (by factors of 9.1 and 4.3, respectively) beyond the values presented in Figure 7 in response to increases in connectivity.

## 4.0 Discussion

The potential "reach" of $MgB_2$ is subject to: (i) The optimization of $H_{c2}$ and $J_c$ through selected dopant additions, and (ii) increasing connectivity and minimizing porosity.

### *4.1 Intrinsic Properties*

***4.1.1. Electronic properties:*** Measurements on doped ("dirty") thin films at liquid He temperatures have demonstrated the attainment of $H_{c20}$s approaching the BCS paramagnetic limit of $H_p$ = 65 T (e.g. [18][19]. Furthermore, given that BCS underestimates $H_p$ by the factor (1+$\lambda$), values as high as 130 T may in principal be achievable [18]. The spectacular thin-film results have inspired researchers to seek equally high $H_{c20}$s in bulk pellets and wires, to achieve which suitable dopants need to be found and suitably dispersed. Likewise the high $J_c$s encountered in some thin film samples have provided goals for $J_c$ development in wires. Progress has been reported in recent years such that, through the use of carefully selected dopants, wire $J_c$s are already on the threshold of the film values. Dopant selection is critical; they should have at least the following properties: (i) They should be able to be uniformly dispersed at an appropriate size scale. (ii) If unreactive with $MgB_2$ or its precursors, their distribution should be such that they ideally interact with the coherence length of $MgB_2$. (iii) If the dopants react with $MgB_2$ or its precursors their products should either modify the Fermi surface of $MgB_2$ so that the critical fields are increased or the reaction products should follow the criteria established in (i). Table 1 summarizes a survey of the several classes of dopants that have been added to the binary base. Of them C, the only element that has so far been substituted for B, has turned out to be the most potent. Although it has been introduced in many forms – elemental C, C-containing inorganic and organic compounds, the most effective medium has been SiC, which is decomposed into Si+C by Mg prior to the $Mg+2B \rightarrow MgB_2$ reaction. Carbon, which produces strong impurity scattering (hence high normal-state resistivity) in the $\pi$ band (as well as the $\sigma$ band) -- cf. Equation (7) -- is responsible for the upturn in $H_{c2}$ at low temperatures [18]. It is important to note that whereas in dirty $MgB_2$ below $T_c$ both the $\sigma$- and $\pi$-bands support superconductivity (mediated by the electron-phonon coupling constants $\lambda_{\sigma\sigma}$, $\lambda_{\pi\pi}$, $\lambda_{\sigma\pi}$, and $\lambda_{\pi\sigma}$) above $T_c$ strong impurity scattering in the $\pi$-bands shunts normal transport current into the $\sigma$-band [101].

***4.1.2. Lattice and Related Properties:*** Although the $\sigma$- and $\pi$-bands play different roles above and below $T_c$ this difference does not invalidate the use of normal-state resistivity as a measure of superconducting state connectivity. Thus the B-G-based resistivity analysis has provided an effective cross-sectional area for general current transport as a measure of connectivity ($1/F < 1$). As well as quantifying the connectivity of powder-formed $MgB_2$ superconductors the B-G-based resistivity analysis yielded values for the *intragrain* residual resistivity, $\rho_0$ (not just its $F$-enhanced value). The increase in $\rho_0$ that accompanied the substitution of SiC (hence C) into $MgB_2$ (Table 3) indicates increased scattering in the $\sigma$-band, the primary conveyer of normal-state current, in agreement with the results of thermal transport measurements on C-doped $MgB_2$ [102]. Deduced from the results of magnetic measurements [18, 21] (but not observed resistively for the above reason) C-doping also enhances scattering in the π-band and hence elevates the low temperature $H_{c2}$ following Equation (7).. Thus C doping enhances scattering in both the σ *and* the π bands, for which reason the observed increase in $\rho_0$ from 11.3 to 38.6 μΩcm is *indirectly* related to a corresponding increase in $H_{c,4.2K}$ of from 21 to more than 33 T [5].

The B-G analysis also yields resistive Debye temperatures, $\theta_R$.. For our three bulk samples, both undoped and doped (average value 680 K), the values obtained are significantly lower than the deduced $\theta_R$s of a pair several dense samples ($\theta_{av}$. ~ 850 K) and a single crystal (950 K). From this we deduce that in the present set of samples granularity and porosity disturb the phonon spectrum more strongly than sample chemistry. To confirm the resistively measured $\theta_R$ values the lattice specific heats of MB700, MBTi800, and MBSiC700 were measured at temperatures up to 300 K. The results (Figure 9(a)) were in satisfactory agreement with published data including those of Wälti et al. [95] who, in a paper dedicated to the specific heat of $MgB_2$, reported a value of 34 J/mole.K at 200 K, cf. 37.9 J/mole.K from Figure 9(a). The PPMS machine's internal software fitted the collected data to a standard Debye function and provided values of the calorimetric Debye temperatures, $\theta_D$. By so doing it ignored the electronic component of total specific heat which, however, contributed less than about 2% to the lattice specific heat at 300 K. The values of $\theta_D$ so obtained, averaging ~670 K, agreed rather well with the resistive results. Wälti et al obtained a $\theta_D$ of 750 K on a pressed and sintered pellet of ex-situ $MgB_2$ powder heat treated for 72 h at 500°C, and Bud'ko et al [96] obtained 750±30 K on a pellet prepared in-situ for 2 h at 950°C. It is likely that these differences in $\theta_D$ respond to the differences in sample preparation conditions.

Given a measured $T_c$, e.g. from the electronic specific heat transition depicted in Figure 9b and either the measured transport- or calorimetric Debye temperature, $\theta_R$ or $\theta_D$, an electron-phonon coupling constant, $\lambda$, can be directly obtained via the following modified version of the McMillan expression [103,104]

$$T_c = \frac{\theta_D}{1.45}\exp\left[\frac{-1.04(1+\lambda)}{(\lambda-\mu^*(1+0.62\lambda)}\right] \tag{13}$$

In selecting a value for $\mu$*, the Coulomb repulsion parameter or "pseudopotential", a survey of the literature ([105-112], but see in particular [111]) yielded $\mu^* = 0.10 \pm 0.02$ (actually equal to the 0.1 estimated for most metals [111]). Presented in Table 4, the results of our analysis provided $\lambda = 0.98$ for the sintered binary pellet and 0.89 and 1.02 for the doped ones. The value for our binary $MgB_2$ lies within the range of reported values [101][105-110][112-114], viz. 0.84 ± 0.21.

What emerges from the specific heat experiment is a global average coupling parameter, $\lambda$, which according to [115] is related to the intraband and interband components, $\lambda_{\sigma\sigma}$, $\lambda_{\pi\pi}$, $\lambda_{\sigma\pi}$, and $\lambda_{\pi\sigma}$, by:

$$\lambda = \left\{\frac{\lambda_{\sigma\sigma}+\lambda_{\pi\pi}}{2}\right\} + \sqrt{\left\{\frac{\lambda_{\sigma\sigma}-\lambda_{\pi\pi}}{2}\right\}^2 + \lambda_{\sigma\pi}\lambda_{\pi\sigma}} \tag{14}$$

It is widely recognized ( e.g. [106][113]) that electron coupling to the so-called $E_{2g}$ phonon mode that represents in-plane B-B bond-stretching oscillations (hence $\lambda_{\sigma\sigma}$) is the dominant mechanism of superconductivity in $MgB_2$. Taken at face value Equation (14) indicates that the non-vanishing of interband coupling permits $\lambda$ to be a function of both $\lambda_{\sigma\sigma}$, *and* $\lambda_{\pi\pi}$.as expected since the σ and π bands contribute to the superconductivity. Values for the individual $\lambda$ components (both transport-related and superconductivity-related) have been tabulated by Mazin et al. [101]

according to whom the superconductive components are: $\lambda_{\sigma\sigma}$ = 1.02, $\lambda_{\pi\pi}$= 0.45, $\lambda_{\sigma\pi}$ = 0.21 and $\lambda_{\pi\sigma}$ = 0.15. Inserting these values into Equation (14) yields a $<\lambda>$ of 1.07 which is very close to $\lambda_{\sigma\sigma}$ itself (cf. our measured 0.98, Table 4) emphasizing the dominance of the in-plane coupling.

### *4.2 Extrinsic Properties*

In general, current transport in all kinds of polycrystalline $MgB_2$ samples tends to be partially blocked by pores and grain-boundary precipitates such that the effective transport cross-sectional area is less than the conductor's geometrical cross-sectional area by a large factor, $F > 1$. Based on the assumption that both superconductive- and normal-state electrical current transport are moderated by the same effective cross-sectional area the result of a normal-state electrical resistivity temperature-dependence, $\rho_m(T)$, measurement can be manipulated so as to provide a value for $F$; or what is the same thing, a "percent connectivity" $100/F$. This quantity is important to the development of practical $MgB_2$ conductors since it leads to estimates of the projected $J_c$s of microstructurally optimized materials quite apart from any improvements that might be expected to accompany the introduction of $H_{c2}$-enhancing and flux-pinning dopants. In Rowell's seminal articles on the subject [7,8], $F$-values for fairly clean $MgB_2$ were derived on the assumption of a universal ideal resistivity $\rho_i(T)$ for all samples.

We have confirmed the validity of this approach for a pair of binary $MgB_2$ samples but have gone on to show that it may break down for heavily doped ones. The breakdown has to do with a sample-to-sample variation of Debye temperature, $\theta_D$. In order to take this into account we have replaced the invariant $\rho_i(T)$ with its Bloch-Grüneisen (B-G) counterpart. As a result excellent fits were obtained for the $\rho_m(T)$s of three in-house-prepared samples as well as five published sets of $\rho_m(T)$ data, as reported elsewhere [116]. The deduced values of $F$ for our bulk powder-processed samples yielded connectivities of from 11 to 23%. Comparable results are expected for in-situ PIT-processed wires indicating that increases in their $J_c$s by factors of 4.3 – 9.1 would accompany the establishment of full connectivity.

It is important to note that $F_p$ curves reported to date combine the effects of flux pinning and connectivity, the material's real $F_p$ being obscured by deficient connectivity. While it is useful to compare the $J_c$ values of various classes of superconductor (e.g., $MgB_2$, YBCO, and $Nb_3Sn$), it is essential to de-convolute the connectivity from the $F_p$. Doing so enables us to define the essential properties in need of improvement. Figure 14 compares the 4-K $F_p$s of an $MgB_2$ strand (actual and projected) with that of an "internal-Sn" $Nb_3Sn$ strand. As a starting point we compare the experimental $F_p$ for the $MgB_2$ (based on the $I_c$ of the powder core divided by the internal cross-sectional area of the sheath or stabilizer) with that of the $Nb_3Sn$ strand (based likewise on its "non-Cu $J_c$"). Clearly $MgB_2$ cannot compete with $Nb_3Sn$ within the liquid-He temperature range depicted here.

Next, let us for a moment assume that a well connected $MgB_2$ strand could be made, such that (based on the bulk-material data of Table 3) its $F_p$ is enabled to increase by factors of 4.3 or 9.1. As illustrated in Figure 14 its 4 K $F_{pmax}$ would increase into the range of 41 to 87 $GN/m^3$. The latter value is comparable to the 90 $GN/m^3$ predicted by Matsushita for a fully connected $MgB_2$ bulk sample based on an alternative site-percolation approach to current transport through a porous medium [10][117]. Not shown in the figure are the even greater enhancements of overall $F_p$ that would result from: (i) increases in $H_{c2}$ and the associated $H_{irr}$ that would follow the addition of suitably chosen dopants especially if they were homogeneously distributed or (ii) the shift to higher fields of the position of the $F_{pmax}$ in response to the inclusion of point pinning centers.

In displaying Figure 14, we in no way aim to minimize the difficulty of achieving the projected $F_p$ goals, or even assert their inevitable attainment. Indeed, while we may surely expect improvements in both intrinsic and extrinsic properties, the degree to which we will be able to approach full connectivity is at present unknown. Similarly, we are constrained in the level of achievable homogeneity in $H_{c2}$ not only by the need to dope uniformly, but also to reduce the anisotropy, or to texture the material. The main point of Figure 14 is to draw attention to the properties most in need of improvement. While flux pinning in $MgB_2$ is already relatively strong (comparable to that of $Nb_3Sn$) connectivity and doping homogeneity are the properties most in need of improvement and control.

### *5. Conclusion*

The prospects for improving the properties of $MgB_2$ in wire form and rendering it suitable for numerous practical applications have been discussed from two essential standpoints, one dealing with the properties of the $MgB_2$ crystal or grain (the intrinsic properties) and the other having to do with ensuring that the intragranular properties are replicated in the material in strand form (the extrinsic aspect of the problem). Regarding intrinsic properties, published thin-film results indicate that the true potential of $MgB_2$ can be approached by the inclusion in it of carefully selected dopants that: (i) increase $H_{c2}$ and associated $H_{irr}$, (ii) increase the volume pinning strength, $F_p$, and especially shifting its maximum to higher reduced fields by augmenting the present grain-boundary pinning with point pinning, (iii) and associate the foregoing with attention to dopant homogeneity, texturing (or, alternatively, anisotropy reduction). But in addition to the incremental improvements that are already being made to the intrinsic properties, spectacular increases in $J_c$ (hence $F_p$) over all fields and temperatures will accompany reductions in porosity and increases in connectivity. Porosity in in-situ and ex-situ-processed materials has been discussed and a new resistivity-based approach to overall connectivity measurement described. It has been estimated that full densification and the removal of intergranular blocking phases could yield across-the-board increases in $J_c$ of up to an order of magnitude. Thus, while all of the parameters listed above are important and will continue to demand attention, the issue of most pressing urgency is connectivity enhancement.

**Acknowledgements**

The $MgB_2$ strands were supplied by HyperTech Research, Columbus, Ohio, courtesy of M. Tomsic and M. Rindfleisch. The research was supported by the U.S. Dept. of Energy, Division of High Energy Physics, under Grants No. DEFG02-95ER40900, DE-FG02-05ER84363, and DE-FG02-07ER84914.

## List of Tables

Table 1. List of dopants added to $MgB_2$.

Table 2. Elemental data and molar volumes.

Table 3. Results of the BG-based resistivity analysis for binary $MgB_2$ and heavily doped (with$TiB_2$ and SiC) $MgB_2$ pellets.

Table 4. Estimation of an electron-phonon coupling constant based on Equation (13) and the resistively determined Debye temperature, $\theta_R$.

Table 1. List of dopants added to $MgB_2$.

| **Nitrides borides silicides** | **Carbon and carbon inorganics** | **Metal oxides** | **Metallic elements** | **Organic compounds** |
|---|---|---|---|---|
| $Si_3N_4$[46-48] | *C nanotubes*[54-57] | $Dy_2O_3$[69] | *Ti*[75-77] | Sugar[86] |
| *WB*[49] | *Nanodiamond*[57-59] | $HoO_2$[70] | *Zr*[77] | malic acid[87] |
| $ZrB_2$[5,44] | *TiC*[60,61] | $Al_2O_3$[71] | *Mo*[78] | maleic anhydride[41] |
| $TiB_2$[5] | *SiC*[38,42,62-65] | *MgO*[45] | *Fe*[79] | paraffin[88] |
| $NbB_2$[5] | $B_4C$[40,66,67] | $TiO_2$[72] | *Co*[80] | toluene[89] |
| | $Na_2CO_3$[68] | $Pr_6O_{11}$[73] | *Ni*[80] | ethanol[89] |
| $CaB_6$[50] | | $SiO_2$[74] | *Cu*[81] | acetone[89] |
| $WSi_2$,[51,52,53] | | | *Ag*[82] | tartaric acid[90] |
| $ZrSi_2$,[51,52] | | | *Al*[83] | ethyltoluene[91] |
| | | | *Si*[84] | |
| | | | *La*[85] | |

Table 2. Elemental data and molar volumes.

| Element | Density $\rho$, g/cc | At/molar wgts, g | Wgts per mol $MgB_2$, g | Vols per mol $MgB_2$, cc |
|---|---|---|---|---|
| Mg | 1.77 | 24.32 | 24.32 | 13.74 |
| B | 2.34 | 10.82 | 21.64 | 9.25 |
| $MgB_2$ | 2.67 | 45.96 | 45.96 | 17.21 |

Table 3. Results of the BG-based resistivity analysis for binary $MgB_2$ and heavily doped (with$TiB_2$ and SiC) $MgB_2$ pellets.

| Sample | $\rho_0$, Ωcm | $\theta_R$, $K$ | $F$ | % Connectivity 100/$F$ |
|---|---|---|---|---|
| MB700 (binary) | 11.29 | 677 | 4.32 | 23 |
| MBTi800 ($TiB_2$) | 9.49 | 764 | 9.50 | 11 |
| MBSiC700 (SiC) | 38.56 | 593 | 7.39 | 14 |

Table 4. Estimation of an electron-phonon coupling constant based on Equation (13) and the resistively determined Debye temperature, $\theta_R$.

| Sample | Onset $T_c$, K | $\theta_R$, $K$ | $\theta_D$, $K$ | $\lambda$ ($\mu^*$=0.10) |
|---|---|---|---|---|
| MB700 (binary) | 38.0 | 677 | 653 | 0.98 |
| MBTi800 ($TiB_2$) | 36.5 | 764 | 745 | 0.89 |
| MBSiC700 (SiC) | 35.5 | 593 | 600 | 1.02 |

## List of Figures

Fig. 1. Resistively measured upper critical fields of eight $MgB_2$ films, after Ferdeghini *et al.* [19].

Fig. 2. DSC traces showing the decomposition of SiC near 524°C and the onset of $MgB_2$ formation near 615°C. Solid line is Mg only (Mg melting at 650°C), Dash-dot line is mixed Mg and B (complete $MgB_2$ below Mg melting), and dashed line is SiC+ Mg powders (reduction of SiC by Mg at 524°C).

Fig. 3. Increases in the irreversibility field of $MgB_2$ in response to doping with 7.5 mol% $ZrB_2$.

Fig. 4. 4 K Transport $J_c$s of $MgB_2$ films made by various techniques including e-beam evaporation, thermal co-evaporation, hybrid physical-chemical vapor deposition, and pulsed-laser deposition (PLD), both in-situ and ex-situ.

Fig. 5. The $J_c$s of some early commercially produced binary $MgB_2$ wires compared to selected thin film results.

Fig. 6. Enhancements in strand-based transport $J_c$ obtained by (i) increasing the Mg content and (ii) doping the result with fine SiC particles.

Fig. 7. A comparison of thin film results to recent measurements on $MgB_2$ strands with various dopants.

Fig. 8. Normalized bulk pinning force curves for sample of an $MgB_2$ strand doped with 5% SiC (particle size 30 nm).

Fig. 9a. Total low temperature specific heat of $MgB_2$ with 5 mol% SiC and 7.5 mol% $TiB_2$ compared to that of a binary, control $MgB_2$ .

Fig. 9b. Electronic specific heat of $MgB_2$ with SiC and $TiB_2$ additions, as compared to a binary control $MgB_2$ .

Fig. 10. SEM micrograph of an $Mg_{1.15}B_2$ monocore PIT strand after 40min/700$^{o}$C.

Fig. 11. Packing density of $MgB_2$ formed from equi-sized Mg and B particles, Model-(1), and that formed from large Mg particles imbedded in fine B powder, Model-(2)

Fig. 12. Schematic illustration of the combined presence of pores and intergranular film showing general porosity or general lack of connectivity.

Fig. 13. Measured resistivity temperature dependence, $\rho_m(T)$, for binary and doped $MgB_2$ samples.

Fig. 14. Field dependence of bulk pinning force density, $F_p$, for present day $MgB_2$ strand, and estimated strand results assuming 4x and 9x greater connectivity. Results are compared to present day HEP class $Nb_3Sn$ conductor.

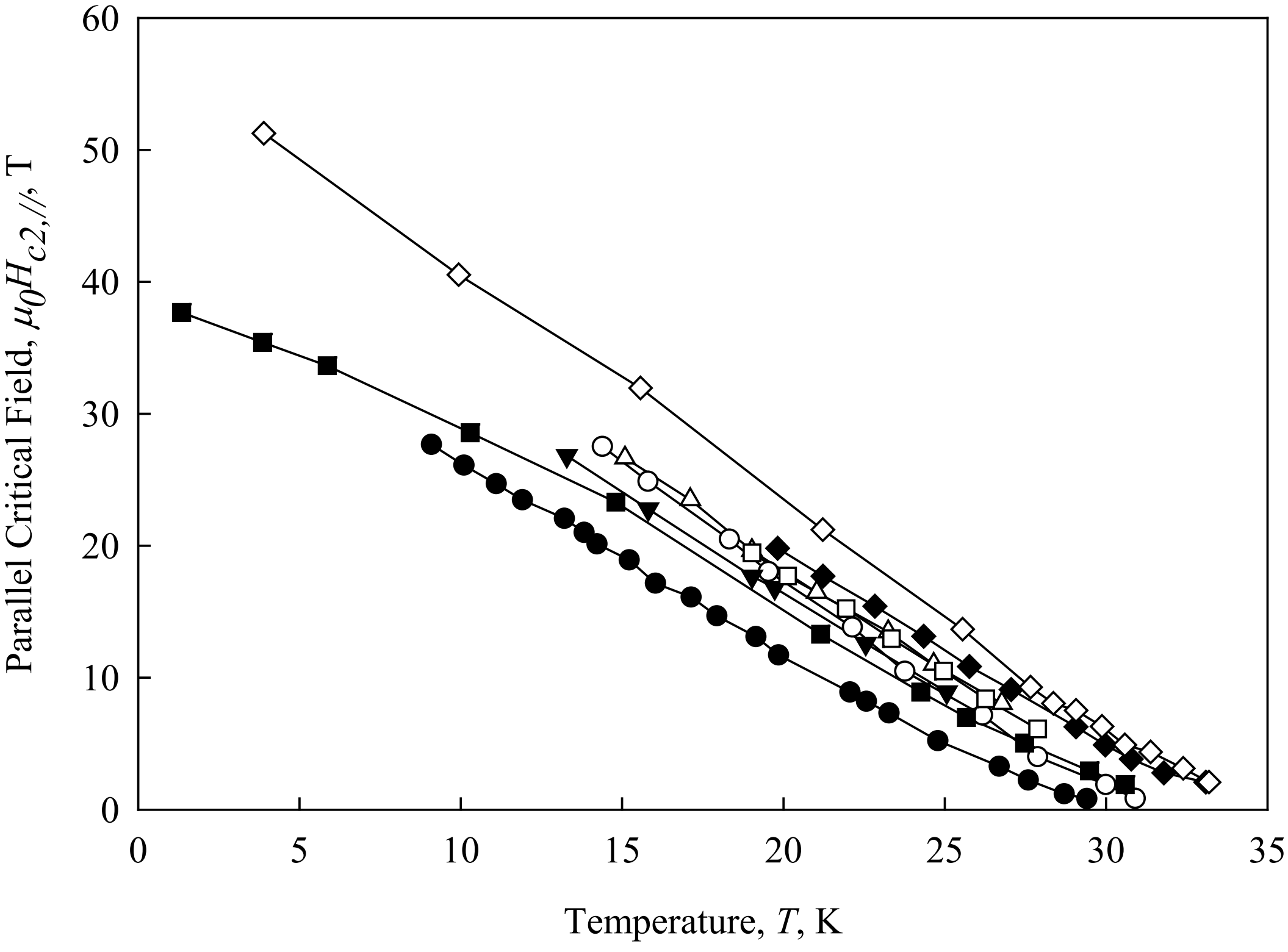

Figure 1.

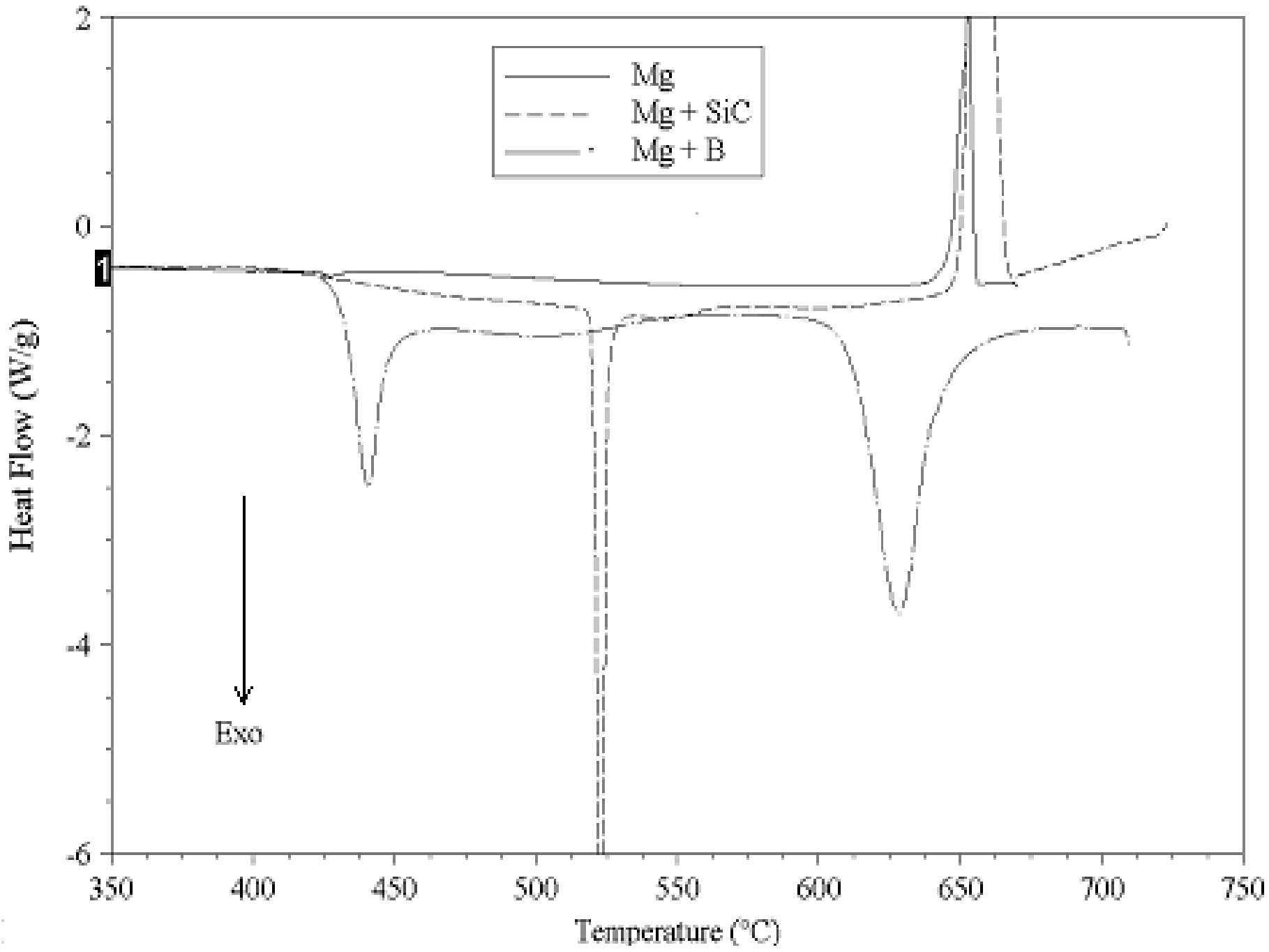

Figure 2.

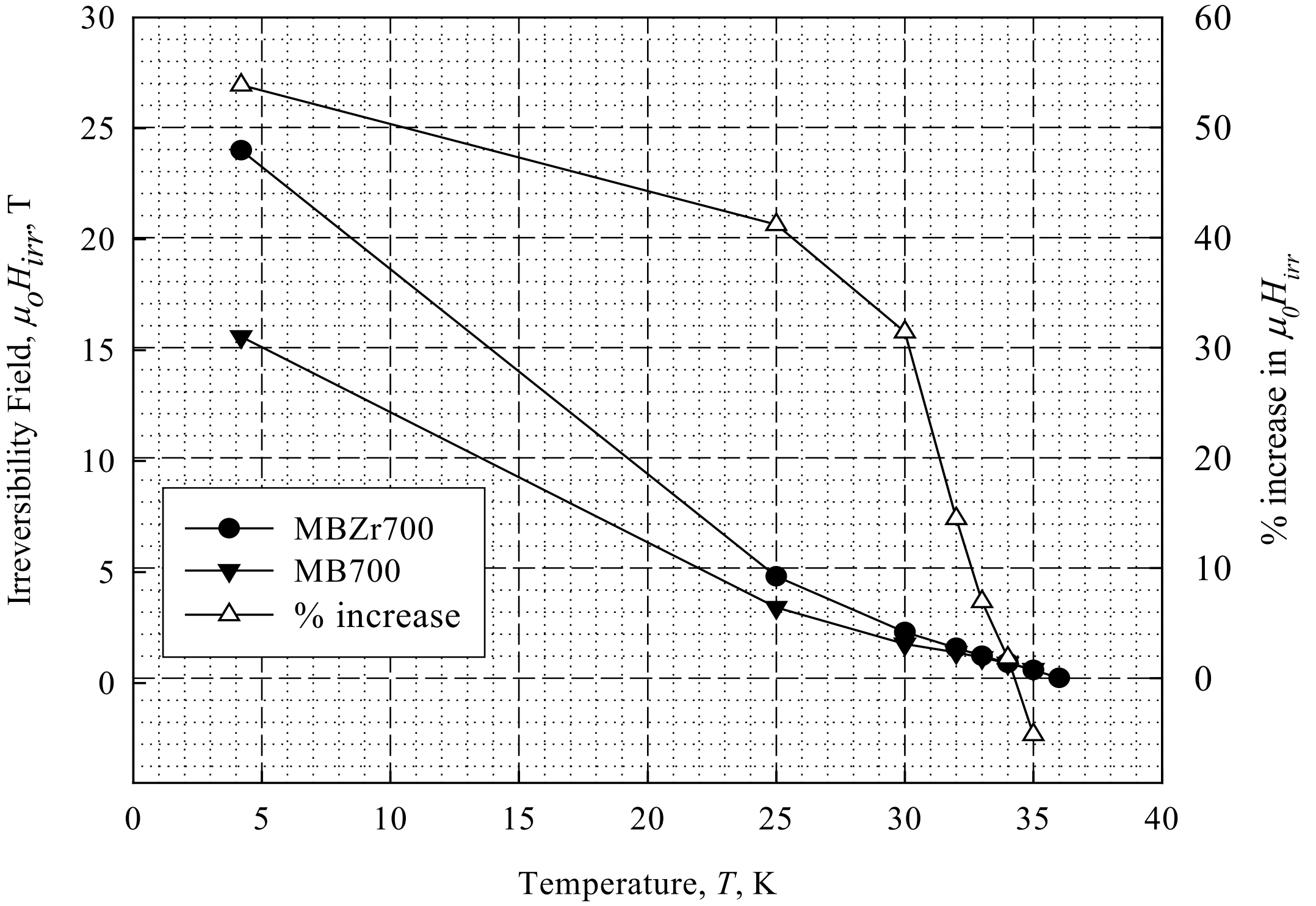

Figure 3.

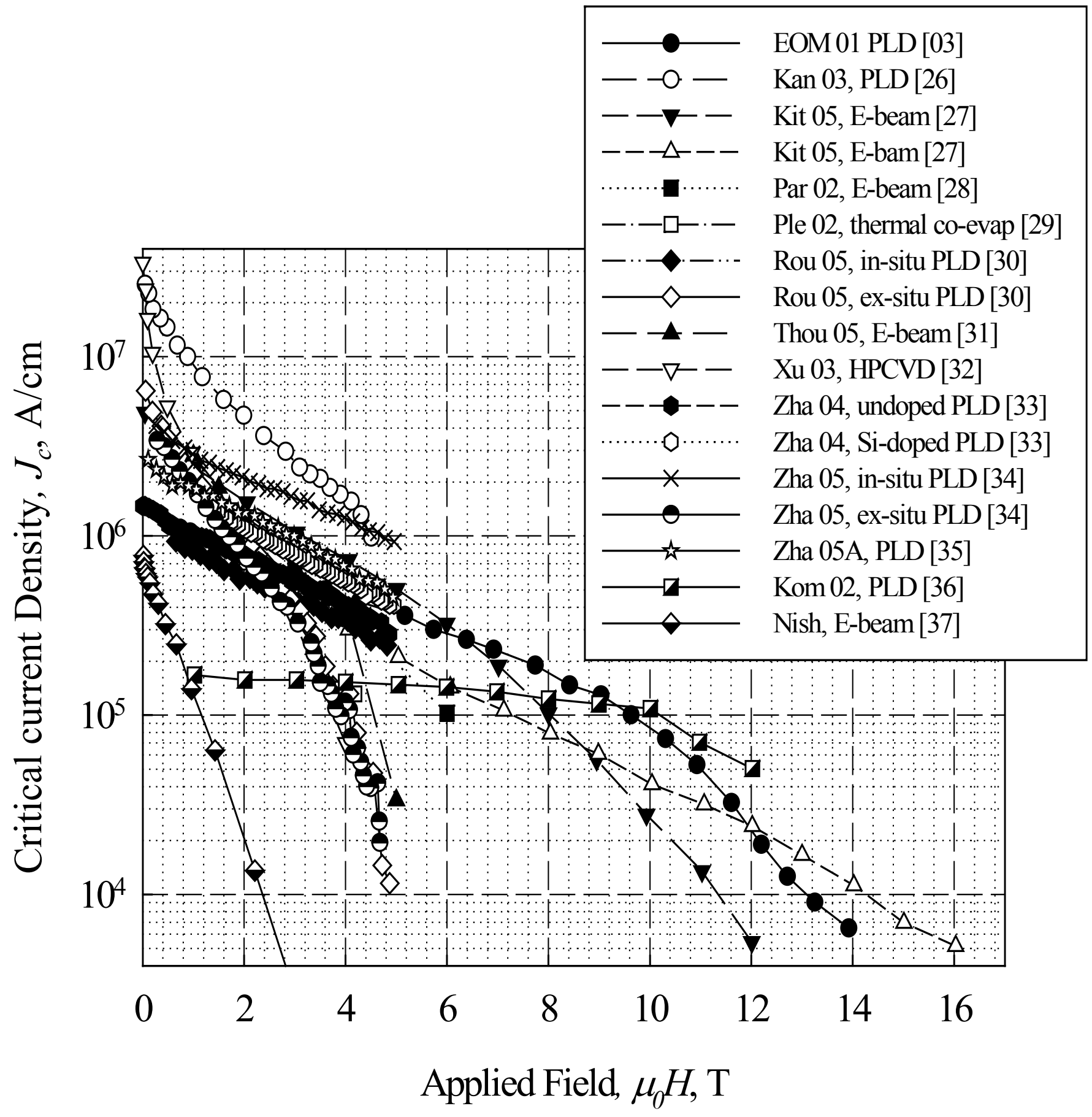
EOM 01 PLD [03]
Kan 03, PLD [26]
Kit 05, E-beam [27]
Kit 05, E-bam [27]
Par 02, E-beam [28]
Ple 02, thermal co-evap [29]
Rou 05, in-situ PLD [30]
Rou 05, ex-situ PLD [30]
Thou 05, E-beam [31]
Xu 03, HPCVD [32]
Zha 04, undoped PLD [33]
Zha 04, Si-doped PLD [33]
Zha 05, in-situ PLD [34]
Zha 05, ex-situ PLD [34]
Zha 05A, PLD [35]
Kom 02, PLD [36]
Nish, E-beam [37]
Critical current Density, $J_c$, A/cm
$10^7$
$10^6$
$10^5$
$10^4$
0
2
4
6
8
10
12
14
16
Applied Field, $\mu_0 H$, T

Figure 4.

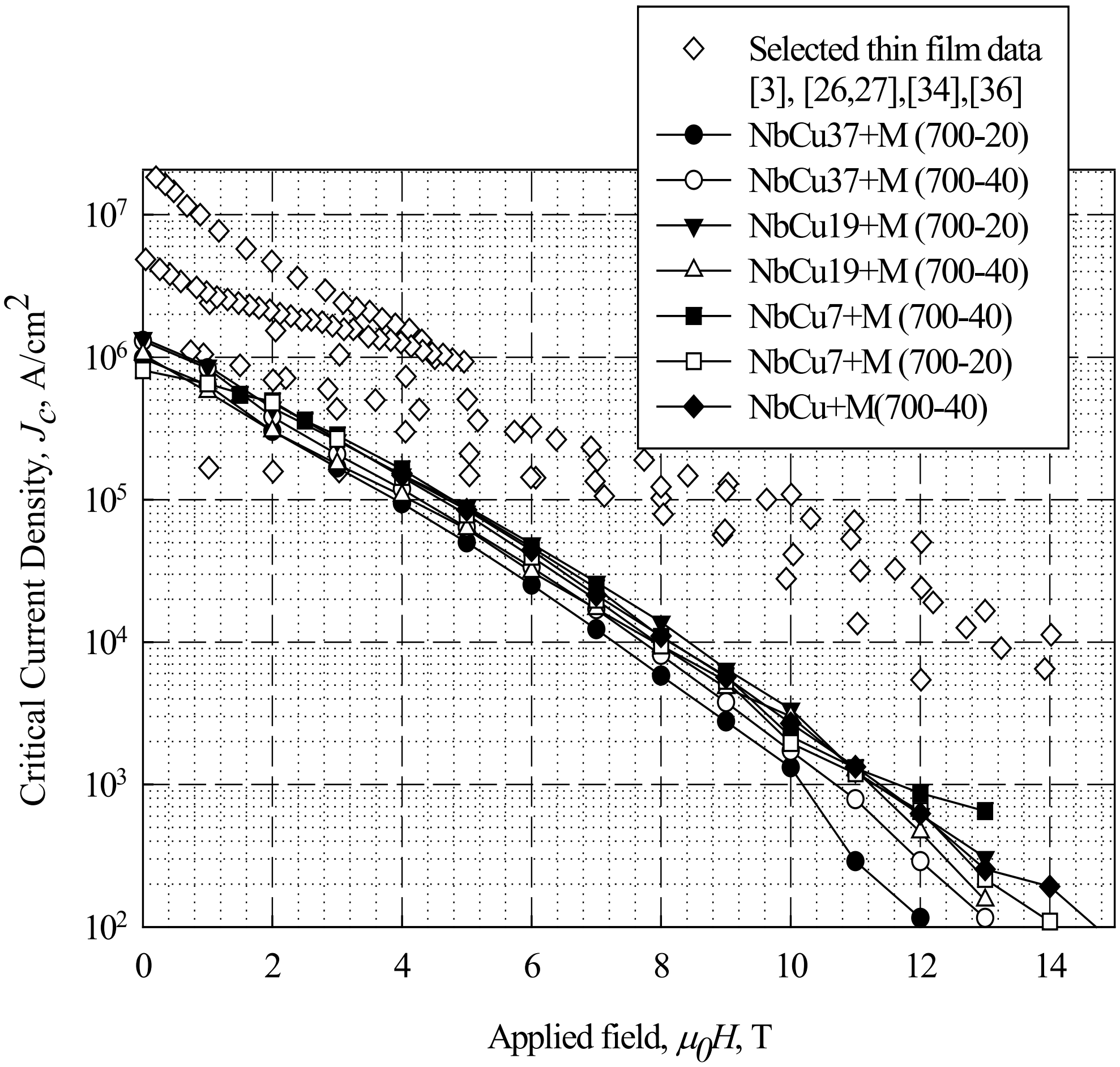

Figure 5.

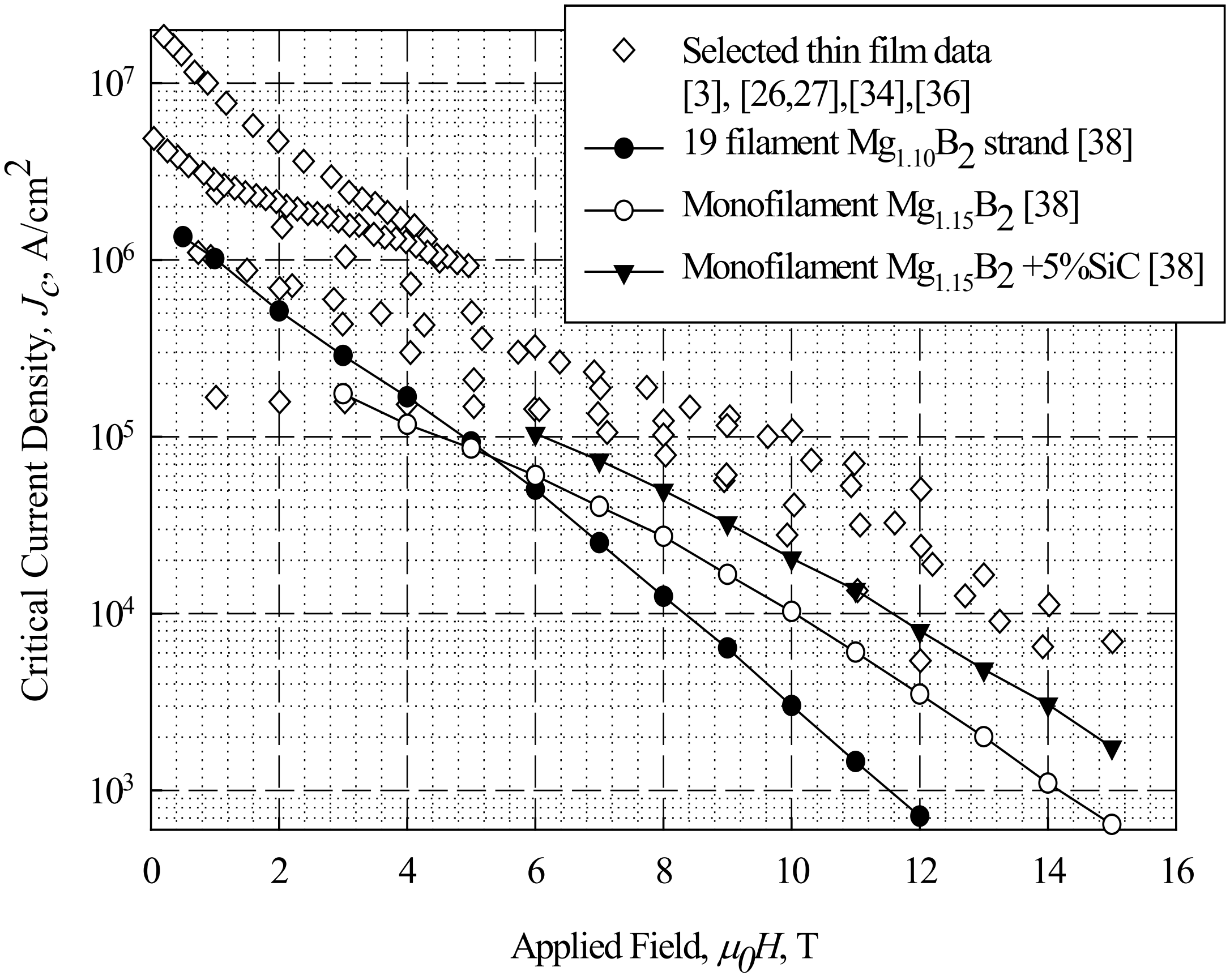

Selected thin film data
[3], [26,27],[34],[36]
19 filament $Mg_{1.10}B_2$ strand [38]
Monofilament $Mg_{1.15}B_2$ [38]
Monofilament $Mg_{1.15}B_2$ +5%SiC [38]
Critical Current Density, $J_C$, A/cm$^2$
$10^7$
$10^6$
$10^5$
$10^4$
$10^3$
0
2
4
6
8
10
12
14
16
Applied Field, $\mu_0 H$, T

Figure 6.

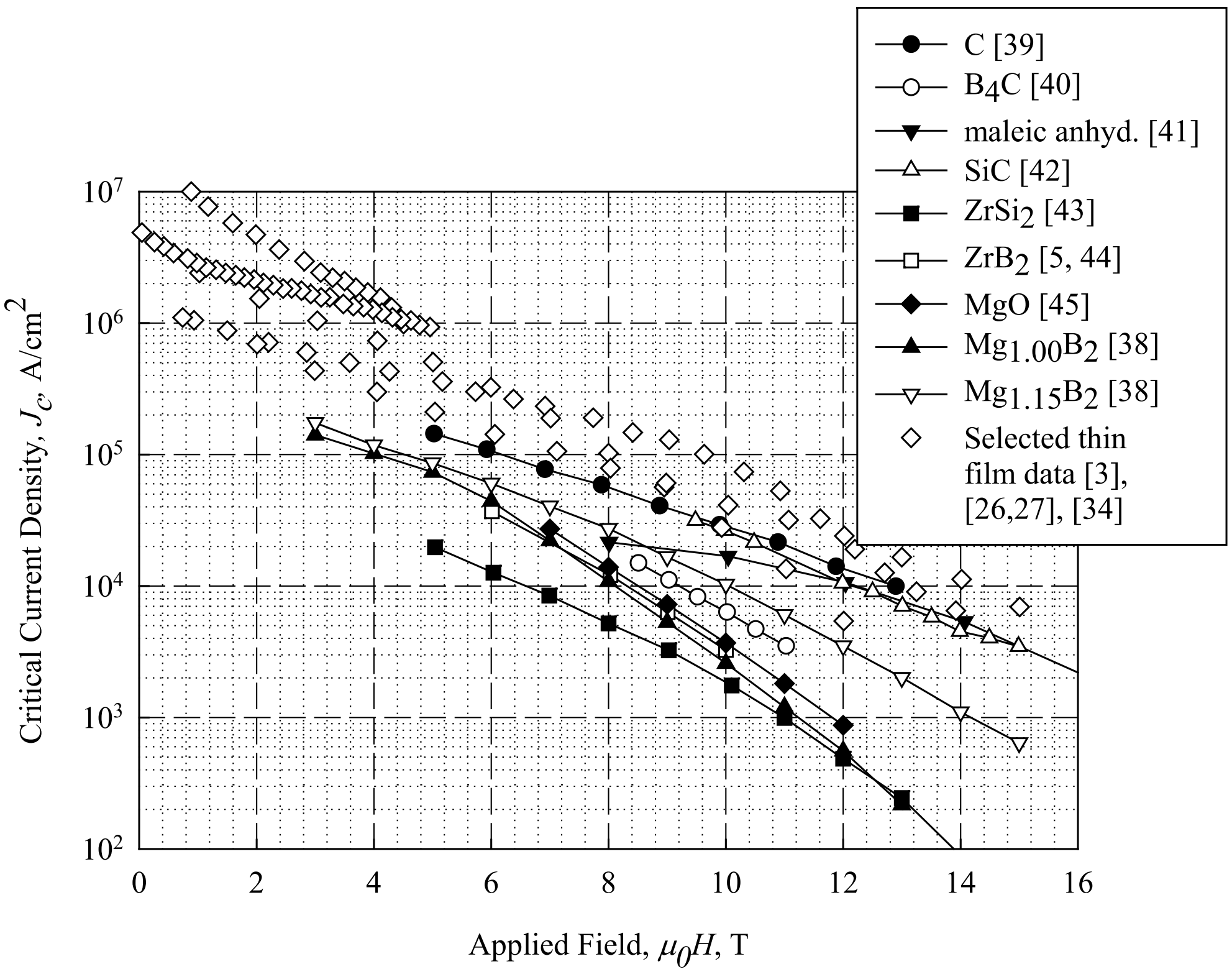

Critical Current Density, $J_c$, A/cm$^2$
Applied Field, $\mu_0 H$, T
C [39]
$B_4C$ [40]
maleic anhyd. [41]
SiC [42]
$ZrSi_2$ [43]
$ZrB_2$ [5, 44]
MgO [45]
$Mg_{1.00}B_2$ [38]
$Mg_{1.15}B_2$ [38]
Selected thin film data [3], [26,27], [34]

Figure 7.

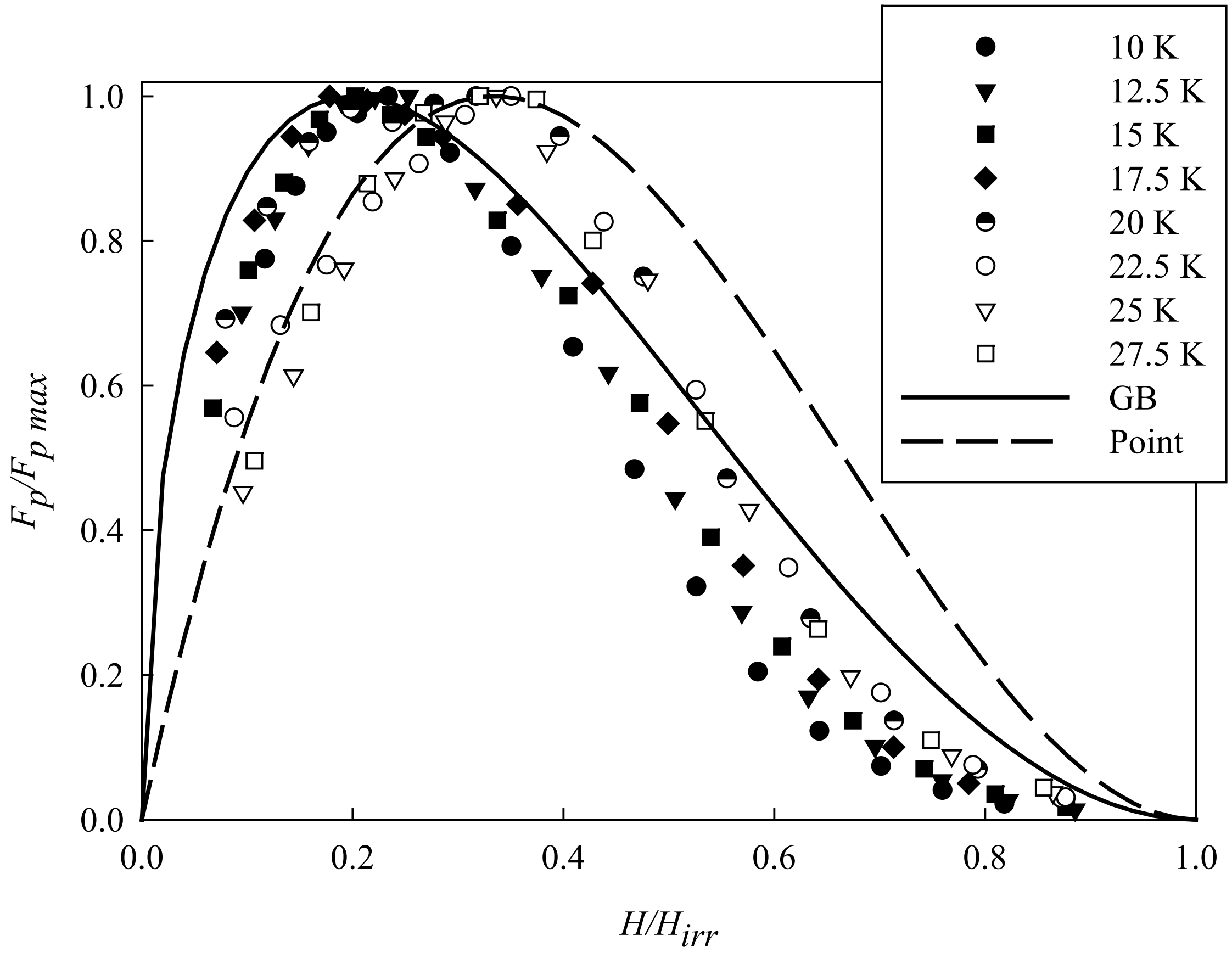

Figure 8.

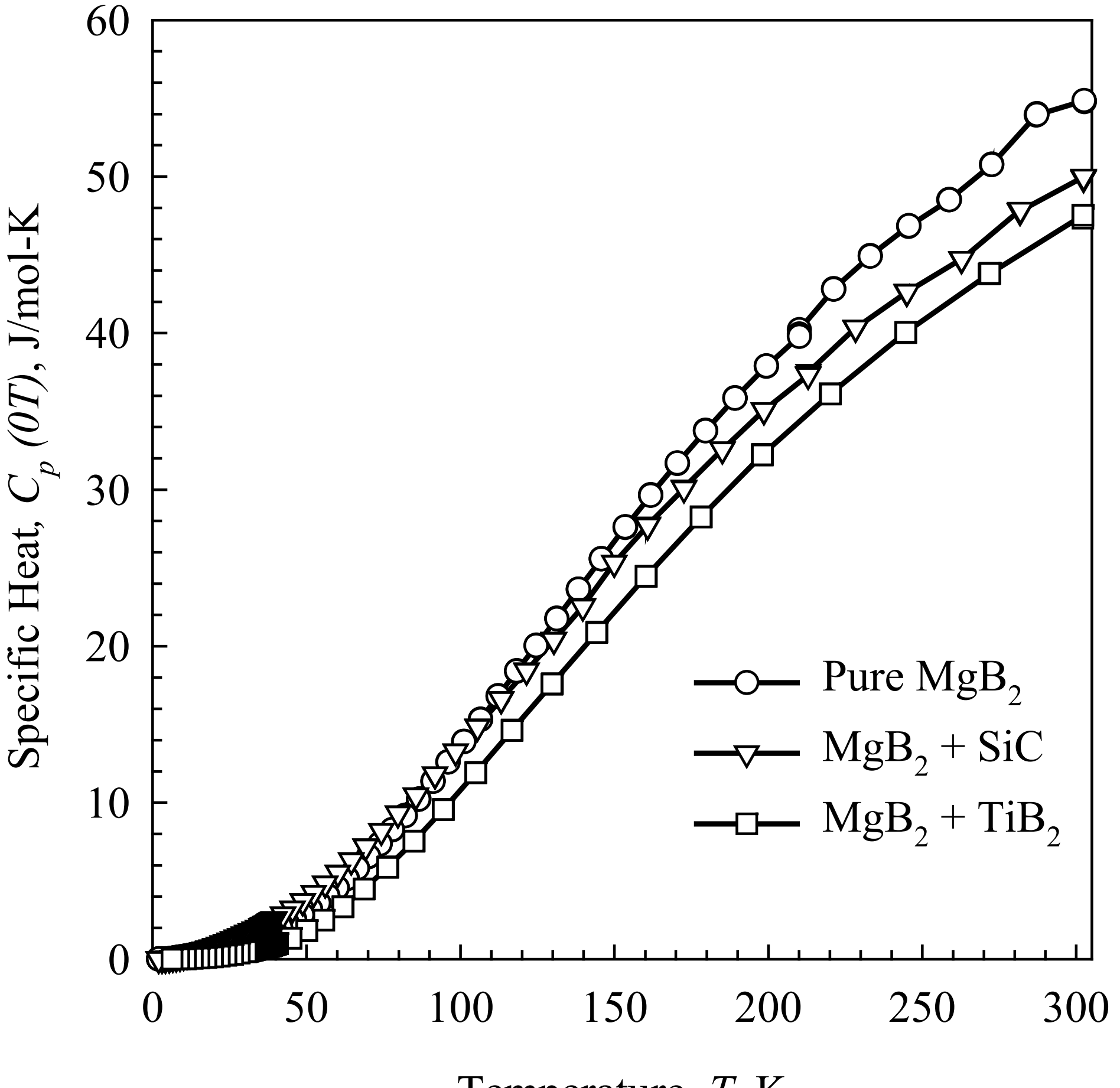

Figure 9a.

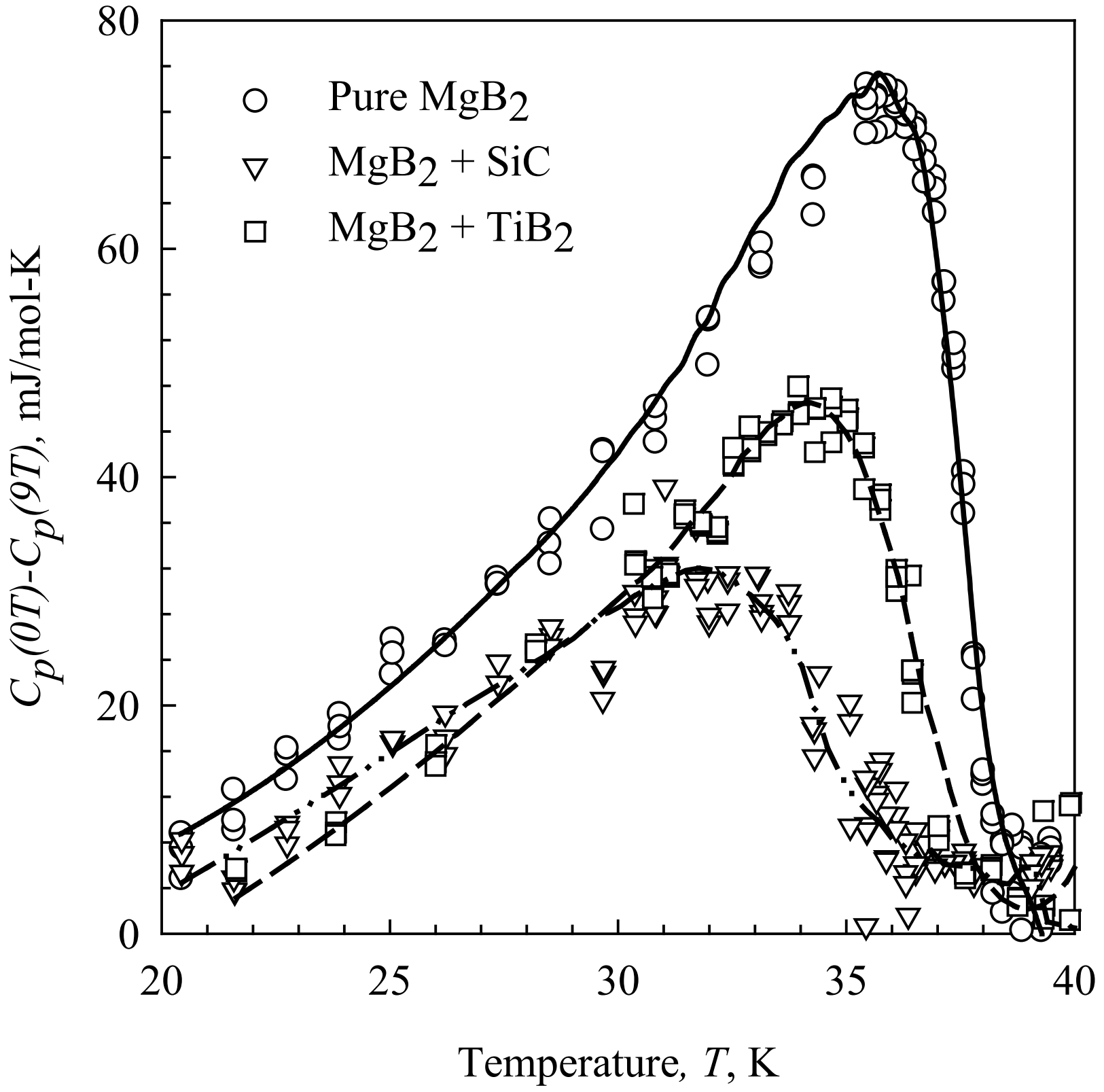

Figure 9b.

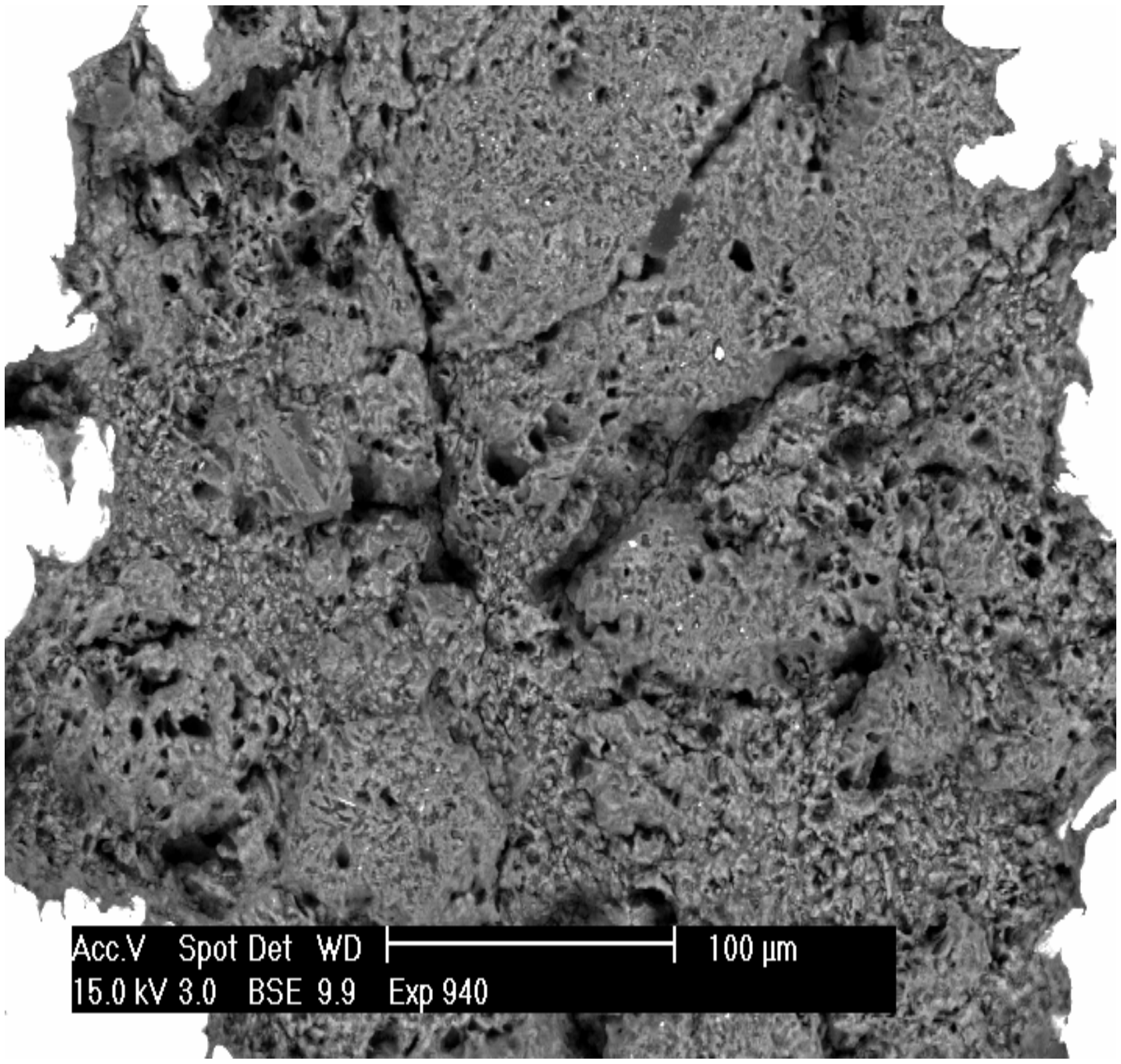

Figure 10.

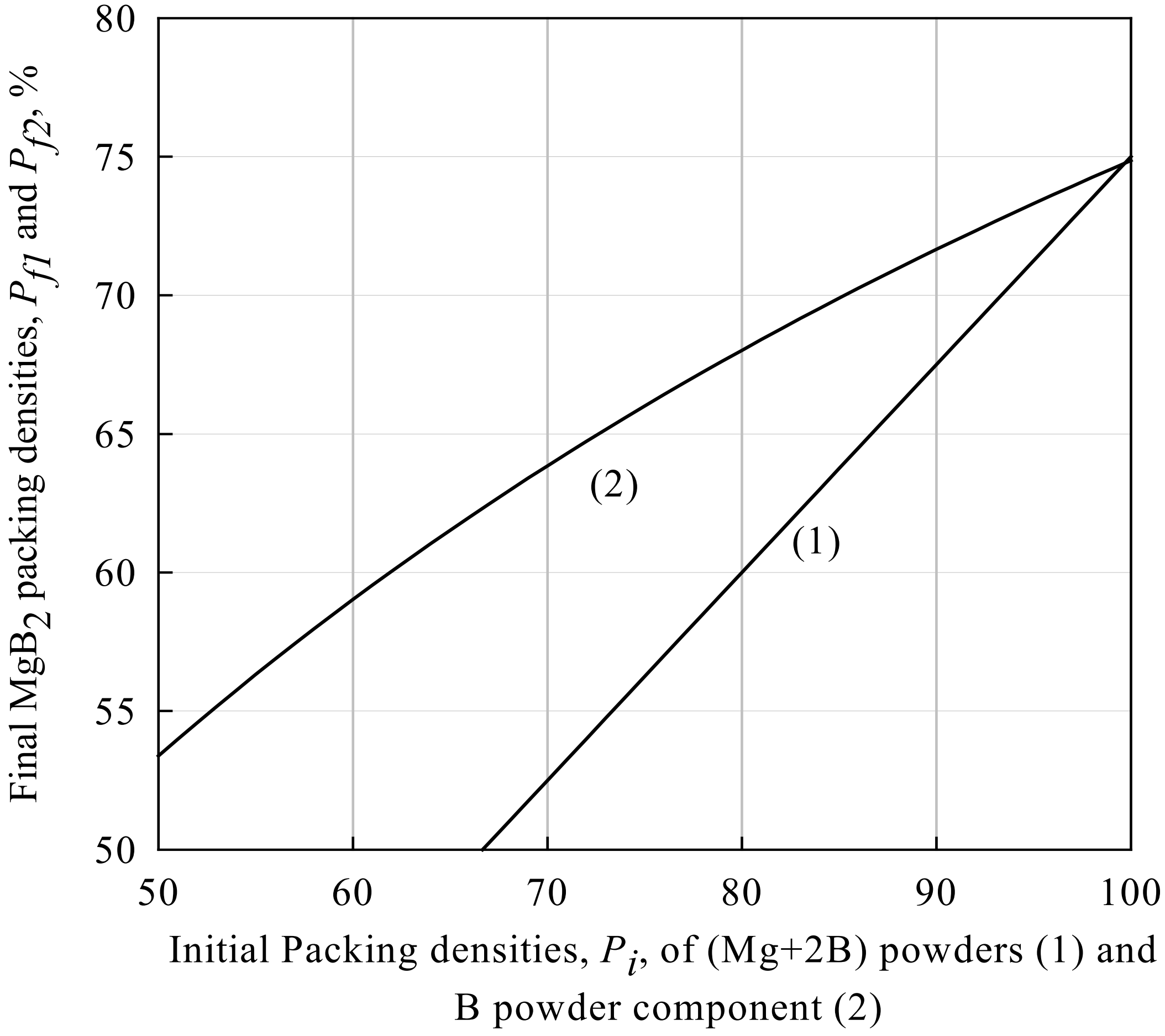

80
75
70
65
60
55
50
50
60
70
80
90
100
(2)
(1)
Final MgB2 packing densities, Pf1 and Pf2, %
Initial Packing densities, Pi, of (Mg+2B) powders (1) and
B powder component (2)

Figure 11.

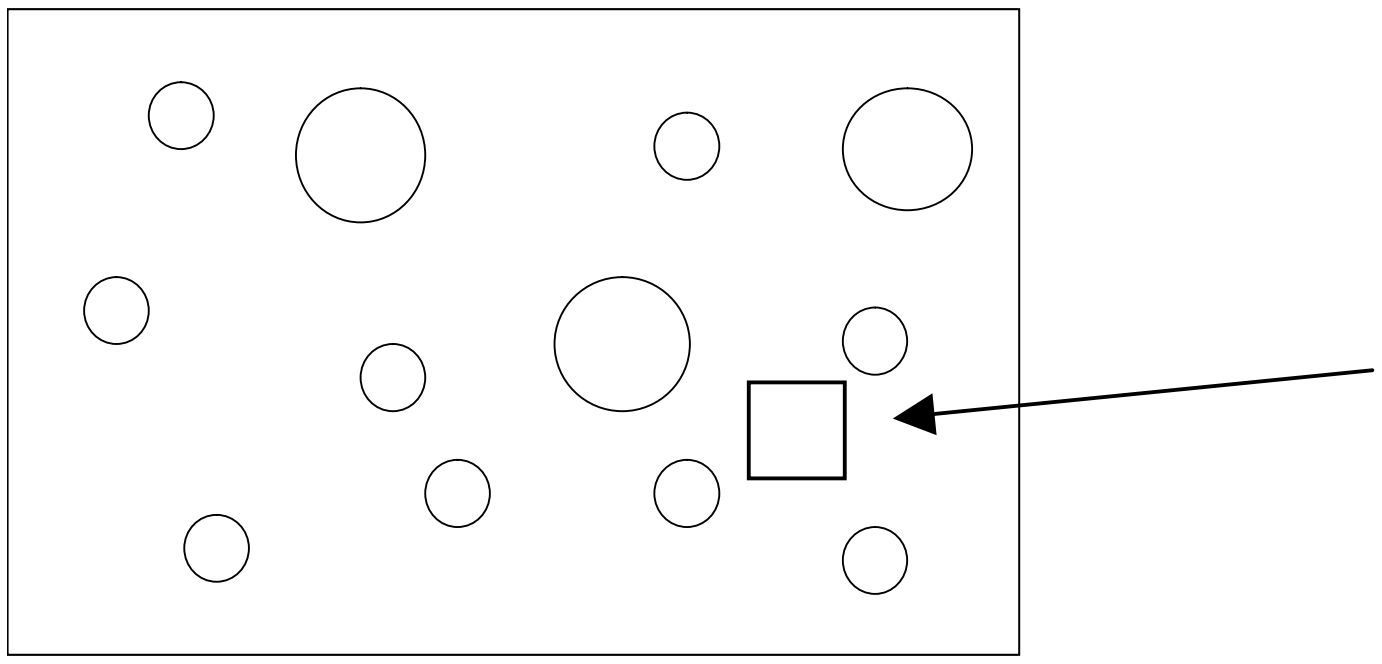

Figure 12.

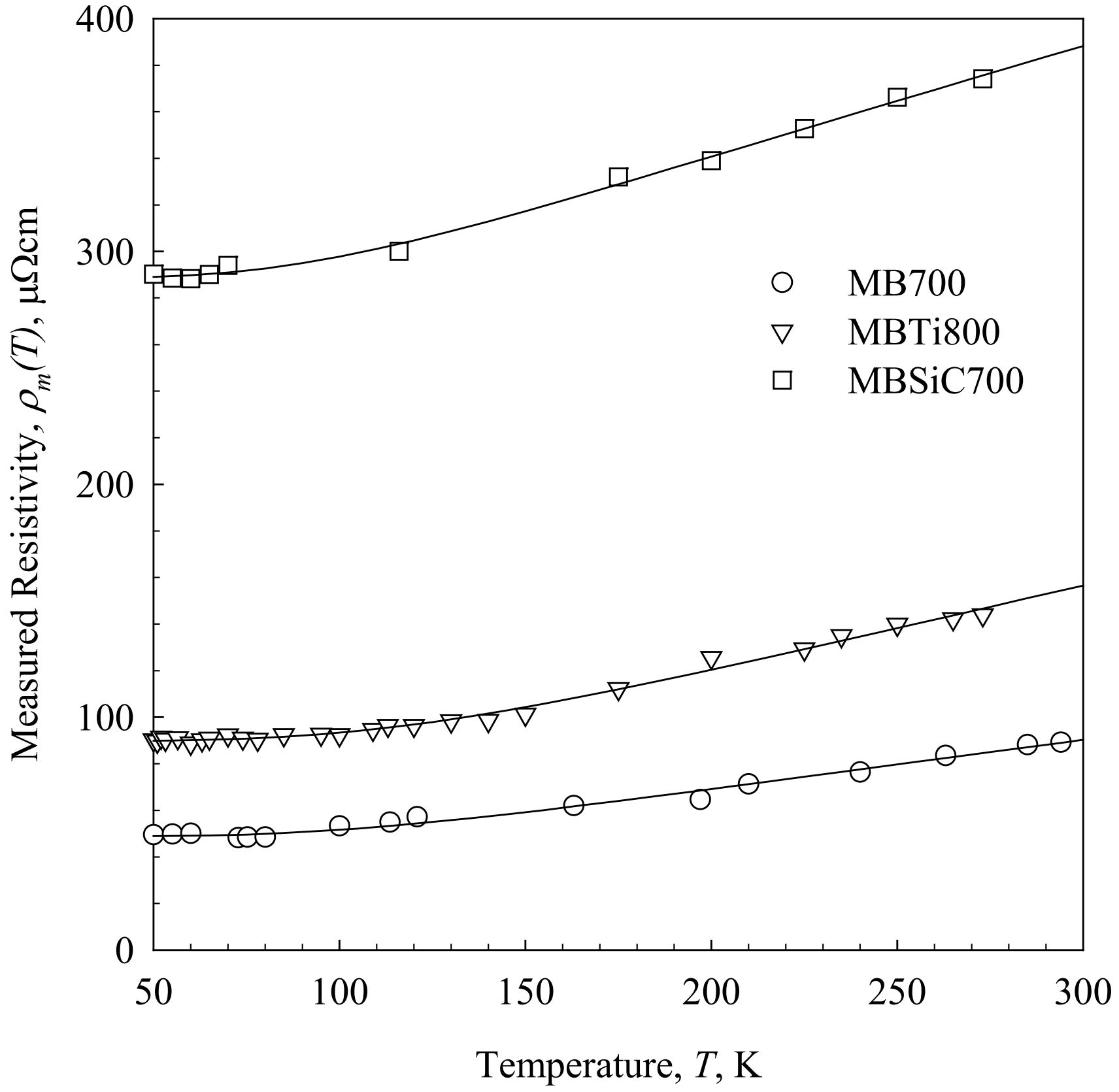

Figure 13.

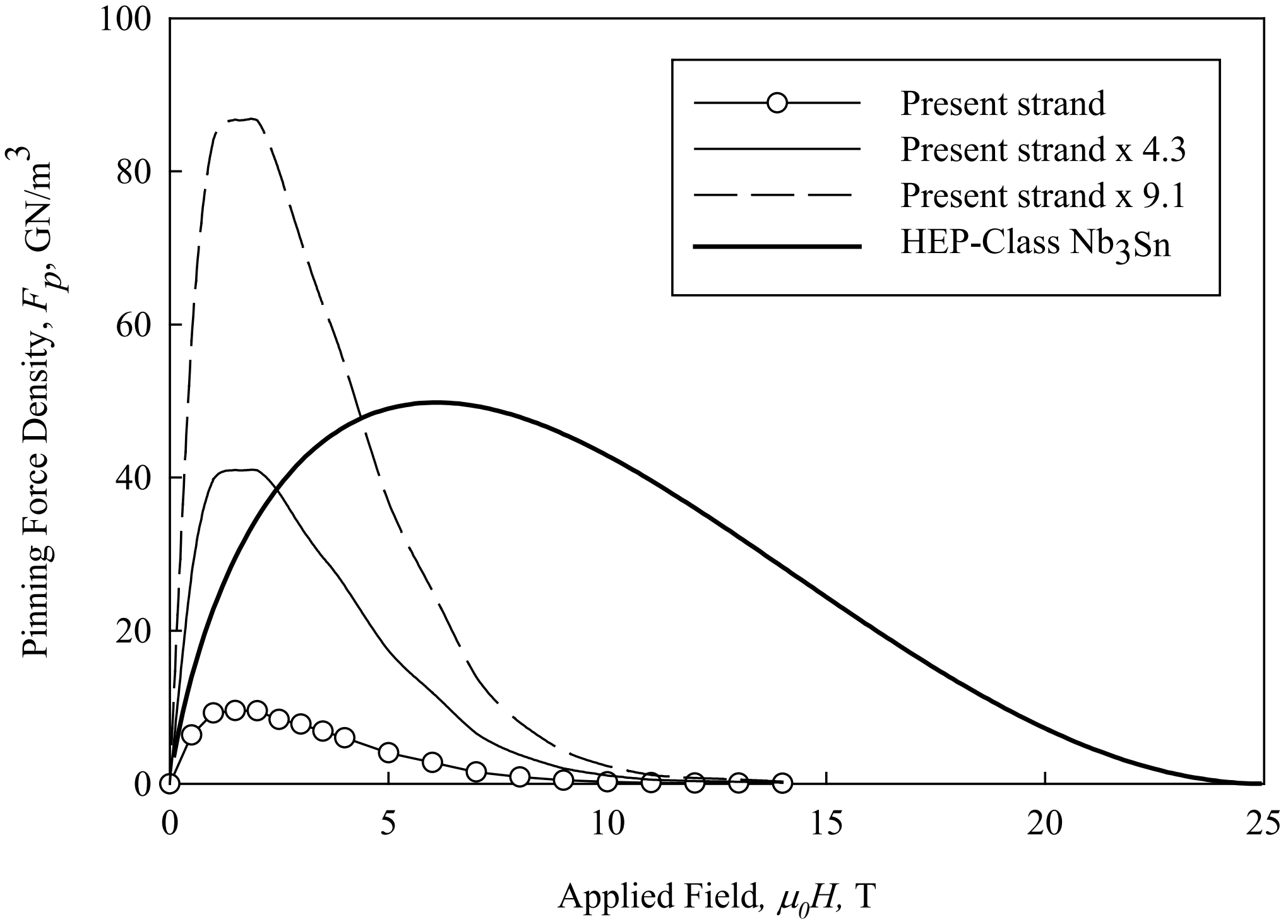

100
80
60
40
20
0
0
5
10
15
20
25
Pinning Force Density, $F_p$, GN/m$^3$
Applied Field, $\mu_0 H$, T
Present strand
Present strand x 4.3
Present strand x 9.1
HEP-Class $Nb_3Sn$

Figure 14